\documentclass[aps,pre,twocolumn,showpacs,superscriptaddress,
amsmath,amssymb,floatfix]{revtex4}
\usepackage{graphicx}
\usepackage{dcolumn}
\usepackage{bm}
\begin{document}
\title{Interfacial structure at a two-dimensional wedge filling transition: 
exact results and a renormalization group study}
\author{J. M. Romero-Enrique}
\affiliation{Department of Mathematics, Imperial College 180 Queen's Gate,
London SW7 2BZ, United Kingdom}
\affiliation{Departamento de F\'{\i}sica At\'omica, Molecular y 
Nuclear, Area de F\'{\i}sica Te\'orica, Universidad de Sevilla, 
Apartado de Correos 1065, 41080 Sevilla, Spain}
\author{A. O. Parry}
\affiliation{Department of Mathematics, Imperial College 180 Queen's Gate,
London SW7 2BZ, United Kingdom}
\author{M. J. Greenall}
\affiliation{Department of Mathematics, Imperial College 180 Queen's Gate,
London SW7 2BZ, United Kingdom}
\begin{abstract}
Interfacial structure and correlation functions near a two-dimensional
(2D) wedge filling transition are studied using effective 
interfacial Hamiltonian models. An exact solution for short range binding 
potentials and results for Kratzer binding potentials 
show that sufficiently close to the filling transition a new  
length scale emerges and controls the decay of the 
interfacial profile relative to the substrate and the correlations between 
interfacial positions above different positions. This new length scale is much 
larger than the intrinsic interfacial correlation length, and it is related 
geometrically to the average value of the interfacial position above the 
wedge midpoint. The interfacial behavior is consistent with a breather mode 
fluctuation picture, which is shown to emerge from an exact decimation 
functional renormalization group scheme that keeps the geometry invariant.
\end{abstract}
\maketitle

\section{Introduction\label{sec1}}

Fluid adsorption in wedge and cone-shaped non-planar geometries has 
attracted much attention in the last few years \cite{Rejmer,Parry1,Parry4,
Parry2,Parry3}. Geometry plays an important role in the surface phase 
diagram, and new phase transitions as the filling transition arise. 
Thermodynamic considerations \cite{Concus,Pomeau,Hauge}
predict that the gas-liquid interface unbinds from the wedge before 
the wetting temperature $T_w$ corresponding to the substrates. So, the 
wedge is completely filled by liquid for temperatures higher than the filling 
temperature $T_f < T_w$, where $T_f$ is given by the condition:
\begin{equation} 
\theta(T_f)=\alpha
\label{fillingcond}
\end{equation}
and $\theta(T)$ is the temperature-dependent contact angle of a liquid 
drop on the planar substrate. Capillary wave models show that the filling 
transition can be critical even though the wetting transition corresponding
to the substrate is first order, and that interfacial fluctuations
are enhanced with respect to the wetting case \cite{Parry4,Parry2}. 
For the 2D wedge filling transition in shallow wedges characterized by an
small angle $\alpha$ respect to the $x$ axis (see below), 
there exists a remarkable covariance relationship between the wedge 
midpoint probability distribution function $P_w^1(l_0)$
in the filling fluctuation regime and the planar $1-$point probability 
distribution function $P_\pi^1(l_0)$ characteristic of a strong-fluctuation 
regime critical wetting transition:
\begin{equation}
P_w^1(l_0;\theta,\alpha)=P_\pi^1(l_0;\theta-\alpha)
\label{covariance}
\end{equation}
where $\theta$ is the contact angle of the liquid droplet on the substrate.
This expression establishes a connection between two apparently unrelated 
phenomena, the deep origin of which is still elusive. 
The covariance relationship has been observed also in acute wedges 
\cite{Abraham1}, Ising model exact calculations \cite{Abraham2} and
computer simulations \cite{Albano}. Although the covariance relationship
is restricted to the interfacial behavior above the wedge midpoint, some 
other quantities like the local susceptibility, which is related to the
$2-$point correlation function, also showed a modified covariance relationship
\cite{Parry3}. Consequently, it is interesting to see if the covariance
extends to higher-order probability distribution functions.

In this paper we study the structure of the interfacial profile
and correlations for 2D wedge filling phenomena. Exact results for the
capillary wave effective Hamiltonian theory in the filling fluctuation
regime are obtained as an extension of the analysis presented
in Ref. \cite{Wood}. The exact results show the appearance of a new 
length scale $\xi_F$ across the wedge close to the critical filling 
transition. This scale controls the decay of the interfacial profile, local
roughness and correlations, and is related geometrically to the wedge 
midpoint average interface position. For the local properties, we found 
a very interesting relationship between the wedge $1-$point probability 
distribution function and the corresponding functions in the planar 
geometry, which can enlighten the origin of the wedge covariance. 

Regarding the two-point correlation functions, we found a confirmation
in the scaling limit of the \emph{breather mode} picture \cite{Parry4,Parry2}, 
which states that the interface is effectively infinitely stiff in the 
filled region and is driven by fluctuations of the wedge midpoint 
interfacial position, i.e. critical effects at 2D wedge filling arise
simply from local translations in the height of the flat, filled interfacial
region.   

Finally, we explain the critical behavior of the filling transition
in the functional renormalization group approach. As the geometry is 
fundamental in the understanding of the critical filling transition, 
we choose a scheme that leaves the wedge geometry invariant. We show that 
the breather mode picture emerges as a straightforward consequence. The 
predictions for the critical behavior are in complete agreement with exact 
solutions.

Our paper is organized as follows. In Section \ref{sec2} 
we describe the continuous transfer matrix formalism and the definition 
of the wedge $n-$point interfacial probability distribution functions. 
We apply this formalism to the case of contact binding potentials in Section 
\ref{sec3} and in particular calculate analytically the $1-$point 
probability distribution function and the $2-$point correlation functions. 
Some results for Kratzer binding potentials will be presented in Section
\ref{sec4}. In Section \ref{sec5} we analyse the breather mode picture and
derive a relation between two important scaling functions. 
Section \ref{sec6} is devoted to the development of a renormalization
group theory of 2D critical filling transition, which requires a 
generalization of previous approaches for critical wetting. 
A brief conclusion is presented in Section \ref{sec7}.  

\section{The formalism \label{sec2}}

\begin{figure}
\includegraphics[width=8.6cm]{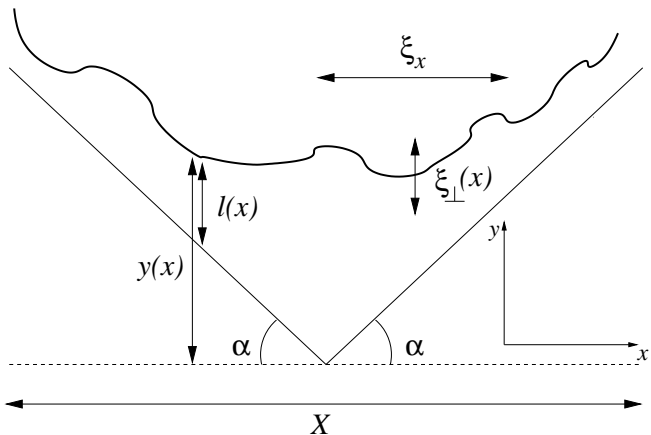}
\caption{Schematic illustration of a typical interfacial configuration 
in the wedge geometry. The relevant correlation length scales $\xi_x$ and
$\xi_\perp(x)$ are also highlighted. Other notation is defined in the 
text.\label{fig1}}
\end{figure}

Consider a two-dimensional wedge formed by the intersection of two equal
planar substrates at angles $\pm \alpha$ to the horizontal (see 
Fig.\ref{fig1}). We suppose that the wedge is in contact with a bulk vapor 
phase at saturation conditions, i.e. in equilibrium with the liquid phase, 
and the substrates preferentially adsorb the liquid phase. 
Our starting point is the effective interfacial Hamiltonian for shallow wedges:
\begin{equation}
\beta H[l]=\int_{-X/2}^{X/2} dx \left\{\frac{\Sigma}{2}\left(\frac{dy}{dx}
\right)^2+W(y(x)-\alpha|x|)\right\}
\label{hameff}
\end{equation}
where $y(x)$ is the interfacial local height respect to the horizontal, $X$ is
the interfacial horizontal length, $k_B T \Sigma$ is the interfacial 
stiffness, $k_B T W(l)$ is the local binding potential and $\beta\equiv 
1/k_B T$. We impose periodic boundary conditions at the ends, 
i.e. $y(-X/2)=y(X/2)$. 

Defining the local relative height between the vapor-liquid interface 
and the substrate $l(x)=y(x)-\alpha|x|$, Eq.(\ref{hameff}) can be 
rewritten as \cite{Parry1}:
\begin{eqnarray}
\beta H[l]&=&X \frac{\Sigma \alpha^2}{2}+ 
\int_{-X/2}^{X/2} dx \Bigg\{\frac{\Sigma}{2}\left(\frac{dl}{dx}
\right)^2\nonumber\\
&+&\Sigma \alpha \left(\frac{dl}{dx}\right)(2H(x)-1) + W(l(x))\Bigg\}
\label{hameff2}
\end{eqnarray}
where $H(x)$ is the Heaviside step function. Integrating by parts to eliminate 
the term proportional to $(dl/dx)$, the effective Hamiltonian can be 
expressed as
\begin{eqnarray}
\beta H[l]&=&
X \frac{\Sigma \alpha^2}{2}+ 2\Sigma \alpha l(X/2)- 2\Sigma \alpha l(0) 
\nonumber\\ &+& \int_{-X/2}^{X/2} dx \left\{\frac{\Sigma}{2}\left(\frac{dl}{dx}
\right)^2+ W(l(x))\right\}
\label{hameff3}
\end{eqnarray}
The first two terms in the equation are irrelevant constants for the 
interfacial properties in the wedge, the third one is the origin of the 
boost factor that decreases the pinning effect of the binding potential 
\cite{Parry1}, 
and the fourth one corresponds to the effective Hamiltonian of an equivalent 
planar interface problem. As the probability distribution of an interfacial 
configuration is proportional to $\exp(-\beta H)$ 
we can relate the wedge and planar probability distributions in a 
straightforward way. In particular, the $n-$point wedge correlation 
functions can be related to $(n+1)-$point correlation functions in the planar 
case by adding the wedge midpoint position. However, the presence of the boost
factor will alter significally the behavior of the wedge correlation functions
with respect to their planar counterparts.  

Our approach is based on a standard application of transfer matrix methods 
\cite{Burkhardt}. The partition function $Z_\pi(l_1,
l_2,x_1,x_2)$ of the interface with fixed endpoints $(x_1,l_1)$ and 
$(x_2,l_2)$ with $x_2>x_1$ in presence of a \emph{planar} substrate is 
defined as the following path integral:
\begin{eqnarray}
Z_\pi(l_1,l_2,x_1,x_2)&\equiv & Z_\pi(l_1,l_2;x_2-x_1)=\nonumber\\
=\int {\cal D}l \exp 
\Bigg(&-& \int_{x_1}^{x_2} dx \left[
\frac{\Sigma}{2} \left(\frac{dl}{dx}\right)^2+W(l)\right]\Bigg)
\label{propagator}
\end{eqnarray}

The partition function Eq. (\ref{propagator}) is the solution of 
the following Schr\"odinger equation:
\begin{equation}
\left[\frac{\partial}{\partial x} + W(l_2) - \frac{1}{2\Sigma}
\frac{\partial^2}{\partial l_2^2}\right]Z_\pi(l_1,l_2;x)=0
\label{schrodinger}
\end{equation}
with the initial condition:
\begin{equation}
Z_\pi(l_1,l_2;0)=\delta(l_2-l_1)
\label{boundary}
\end{equation}
where $\delta(x)$ is the Dirac delta function. Formally, the partition 
function can be expressed as:
\begin{equation}
Z_\pi(l_1,l_2;x)= \sum_i \psi^*_i(l_1)\psi_i(l_2)\exp(-E_i x)
\label{spectral}
\end{equation}
where $\psi(l)$ and $E_i$ are the eigenfunctions and eigenvalues of the 
time-independent Schr\"odinger equation:
\begin{equation}
- \frac{1}{2 \Sigma} \psi_n^{''}(l)+W(l)\psi_n(l)=E_n\psi_n(l)
\label{schrodinger2}
\end{equation}
with appropriate boundary conditions. In the thermodynamic limit 
$Z_\pi \sim \exp(-\beta f X)$ as $X\to \infty$, where 
$\beta f=\Sigma (\cos \theta -1)$ is the excess free energy per interfacial 
length. Consequently, Eq. (\ref{spectral}) implies that $E_0=\beta f$, so
that in the low contact angle limit, $E_0\approx -\Sigma \theta^2/2$. 

The $n$-point distribution functions can be obtained in terms of 
$Z_\pi(l_1,l_2;x)$ as:
\begin{eqnarray}
P_\pi(1;\ldots;n)=\lim_{X\to \infty} 
\frac{\prod_{i=0}^{n} Z_\pi(l_i,l_{i+1}; x_{i+1}-x_i)}
{Z_\pi(l_{-X/2},l_{X/2};X)}
\nonumber\\ = \psi_0(l_1) \psi_0^*(l_n)
\prod_{i=1}^{n-1}Z_\pi(l_i,l_{i+1};x_{i+1}-x_i)
\textrm{e}^{E_0 (x_{i+1}-x_i)}
\label{corrplanar}
\end{eqnarray}
where $i\equiv (l_i;x_i)$, $x_{n+1}=-x_0\equiv X/2$, and $l_0=l_{n+1}=
l_{X/2}$. For $n=1$, $P_\pi(i)\equiv |\psi_0(l_i)|^2$.
From Eqs. (\ref{corrplanar}) and 
(\ref{spectral}) it is clear that if the distance between two subsets 
$\{x_1,\ldots,x_m\}$ and $\{x_{m+1},\ldots,x_n\}$ is much greater than 
the \emph{planar correlation length} $\xi_\parallel\equiv 1/(E_1-E_0)$ 
(with $E_1$ the first excited state eigenvalue), the distribution function 
factorizes and the two subsets become uncorrelated.
 
The  $n$-point  \emph{wedge} distribution functions 
$P_w(1;\ldots;n)$ can be expressed, in general, 
in terms of $(n+1)-$point planar distribution functions. So, for a 
set $\{x_{-m}<\ldots<x_{-1}<0<x_1<\ldots x_n\}$, they can be expressed as:
\begin{eqnarray}
P_w(-m;\ldots;n)&=&\int_0^\infty dl_0 
\frac{\textrm{e}^{2\Sigma\alpha l_0}}
{\langle 0|\textrm{e}^{2\Sigma\alpha l_0}|0\rangle}
P_\pi(-m;\ldots;\nonumber\\ -1;0;1;\ldots;n)
&=&\frac{P_w(-1;1)P_\pi(-m;\ldots;n)}{P_\pi(-1;1)}
\label{wedgecorr}
\end{eqnarray}
where $\langle n|f(l)|m\rangle \equiv 
\int_0^\infty dl \psi_n(l) f(l) \psi_m^*(l)$. If $0\le x_1<\ldots<x_n$, 
the expression of $P_w(1;\ldots;n)$ is slightly simpler:
\begin{eqnarray}
P_w(1;\ldots;n)&=&\int_0^\infty dl_0 
\frac{\textrm{e}^{2\Sigma\alpha l_0}}
{\langle 0|\textrm{e}^{2\Sigma\alpha l_0}|0\rangle}
P_\pi(0;1;\ldots;n)\nonumber\\
&=&\frac{P_w(1) P_\pi(1;\ldots;n)}{P_\pi(1)}
\label{wedgecorr2}
\end{eqnarray} 
A similar expression is found if $x_1<\ldots<x_n\le 0$.
Finally, if $x=0$ is included in the $x$ set, the wedge 
$n$-point distribution function reduces to:
\begin{equation}
P_w(-m;\ldots;n)= 
\frac{\textrm{e}^{2\Sigma\alpha l_0}}
{\langle 0|\textrm{e}^{2\Sigma\alpha l_0}|0\rangle}
P_\pi(-m;\ldots;n)
\label{wedgecorr3}
\end{equation}

Although this approach is general for arbitrary binding potentials, we 
will restrict ourselves to some special cases. The first case will be
contact potentials, in which $W(l)=0$ 
for $l>0$, $W(l)=+\infty$ for $l<0$ and at the wall the eigenfunctions fulfill
the boundary condition \cite{Burkhardt}:
\begin{equation}
\frac{\partial}{\partial l} \ln \psi(l)\Bigg|_{l=0}=-\tau
\label{boundary2}
\end{equation}
where $\tau$ is proportional to the deviation from the critical wetting 
temperature. For $\tau>0$ the contact angle is related to $\tau$ via 
$\tau= \Sigma \theta$ \cite{Burkhardt}. These potentials can be 
understood as the limiting case of a square-well binding potential when 
the well width tends to zero. Its importance is threefold. First, this case 
corresponds to the filling fluctuation regime, that previous studies show 
to be the relevant one for potentials which decay faster than $1/l$. 
Secondly, there is an analytical expression for $Z_\pi(l_1,l_2;x)$ 
\cite{Burkhardt}, given by:
\begin{eqnarray}
& &Z_\pi(l_1,l_2;x)=\sqrt{\frac{\Sigma}{2\pi x}}\left(\textrm{e}^{-
\frac{\Sigma(l_2-l_1)^2}{2x}}+\textrm{e}^{-\frac{\Sigma(l_1+l_2)^2}
{2x}}\right)
\nonumber\\
&+& \tau \textrm{e}^{\frac{\tau^2 x}{2\Sigma} - \tau(l_1+l_2)} 
\textrm{erfc}\left(\sqrt{\frac{ \Sigma}{2\pi x}}(l_1+l_2)-\tau 
\sqrt{\frac{x}{2\Sigma}}\right)
\label{propagator2}
\end{eqnarray}
Finally, this case can be compared to more microscopic results, like the exact
solutions of the interfacial properties of the corner filling of an Ising 
model. 

Another interesting case is the Kratzer binding potential \cite{Grosche}:
\begin{equation}
W(l)=-\frac{\phi \theta}{l} + \frac{w}{l^2}
\label{kratzer}
\end{equation}
where $\phi=(1+\sqrt{1+8\Sigma w})/2$ and we assume Dirichlet boundary 
conditions at the origin. Previous studies indicate that this class of
binding potentials corresponds to the marginal case between the mean-field and 
fluctuation dominated regimes for the critical filling transition. The 
Laplace transform of $Z_\pi(l_1,l_2;x)$, $\tilde{Z}_\pi(l_1,l_2,E)$, is 
given by \cite{Grosche}:
\begin{eqnarray}
& &\tilde{Z}_\pi(l_1,l_2,E)=\int_0^\infty dx\ 
\textrm{e}^{Ex}Z_\pi(l_1,l_2;x)
\nonumber\\
&=&\frac{\sqrt{\frac{E_0}{E}}\Gamma\left[\phi\left(1-
\sqrt{\frac{E_0}{E}}\right)\right]}{\theta\ \Gamma\left[2\phi\right]}
W_{\phi\sqrt{\frac{E_0}{E}},\phi-1/2}\left(\sqrt{-8\Sigma E}l_>\right)
\nonumber\\
&\times& M_{\phi\sqrt{\frac{E_0}{E}},\phi-1/2}\left(\sqrt{-8\Sigma E}l_<\right)
\label{laplacecoulomb}
\end{eqnarray}
where $E_0=-\Sigma \theta^2/2$, $l_>=\max(l_1,l_2)$, $l_<=\min(l_1,l_2)$,
$\Gamma(x)$ is the gamma function, and finally $M_{\kappa,m}(z)$ and
$W_{\kappa,m}(z)$ are Whittaker functions, related to confluent hypergeometric
functions.
 
\section{Exact results for contact binding potentials \label{sec3}}

In this Section we will obtain and analyze some relevant wedge distribution
functions for contact binding potentials. In particular, we will revisit
the $1-$point distribution function (considered previously by our group 
\cite{Wood}) and the $2-$point height-height correlation function between the 
midpoint and any other interfacial positions. Related quantities as the
average interfacial profile $\langle l(x)\rangle_w$, the local roughness 
$\xi_\perp(x)$ and the correlation length across the wedge $\xi_x$ (see 
Fig.\ref{fig1}) will be also obtained.

Some results are already known for the $1-$point distribution functions. 
The probability distribution function for the midpoint $x=0$ interfacial
height is given by \cite{Parry1}:
\begin{equation}
P_w^1(l_0;\theta,\alpha)\equiv P_w(l_0,0)=2\Sigma 
(\theta-\alpha)\textrm{e}^{-2\Sigma (\theta-\alpha) l_0}
\label{PDF0}
\end{equation}
that verifies the remarkable covariance relationship Eq. (\ref{covariance}).

For arbitrary $x\ge 0$ the $1-$point distribution function has the expression 
\cite{Wood}:
\begin{eqnarray}
P_w(l,x) &=& \Sigma\theta \textrm{e}^{-2\Sigma\theta l}
\textrm{erfc}\left(-\sqrt{\frac{\Sigma x}{2}}\theta
+\sqrt{\frac{\Sigma}{2x}}l\right) \nonumber\\
&+&\Sigma (\theta -\alpha) \textrm{e}^{2\Sigma(\alpha-
\theta)l}\textrm{e}^{2\Sigma \alpha x (\alpha - \theta)}\nonumber\\
&\times&\textrm{erfc}\left(\sqrt{\frac{\Sigma x}{2}}(\theta-2\alpha)-
\sqrt{\frac{\Sigma}{2x}}l\right)\nonumber\\
&-&\Sigma \alpha \textrm{e}^{-2\Sigma\alpha l}
\textrm{e}^{2\Sigma \alpha x (\alpha - \theta)}\nonumber\\
&\times&\textrm{erfc}\left(\sqrt{\frac{ \Sigma x}{2}}(\theta-2\alpha)+
\sqrt{\frac{\Sigma}{2x}}l\right)
\label{PDFx}
\end{eqnarray}
For $x<0$, we have the symmetry $P_w(l,x)=P_w(l,-x)$, so hereafter we will 
consider only the case $x\ge 0$.

The moments $\langle l^n(x)\rangle_w$ can be obtained after some algebra. 
The average interfacial position profile reads:
\begin{eqnarray}
\langle l(x)\rangle_w = \frac{1}{2\Sigma \theta}+
\sqrt{\frac{x}{2\pi\Sigma}} \textrm{e}^{-\frac{\Sigma \theta^2}{2}x}
\nonumber\\+
\left[\frac{\theta}{\theta-\alpha}-\frac{\theta}{\alpha}\right]
\frac{\textrm{e}^{2\Sigma \alpha x (\alpha - \theta)}}{4\Sigma\theta}
\textrm{erfc}\left(\sqrt{\frac{ \Sigma x}{2}}
(\theta-2\alpha)\right)\nonumber\\
+\left[\frac{1}{4\Sigma\theta}\left(\frac{\theta}{\theta-\alpha}+
\frac{\theta}{\alpha}-2\right)-\frac{\theta x}{2} \right]
\textrm{erfc}\left(\sqrt{\frac{ \Sigma \theta^2}{2}x}
\right)
\label{av1}
\end{eqnarray}
The wedge excess adsorption $\Gamma_w$ measured with respect to the planar 
case can be obtained as:
\begin{eqnarray}
\Gamma_w=2(\rho_l-\rho_g)\int_0^\infty \left(\langle l(x) \rangle_w-
\frac{1}{2\Sigma\theta}\right)dx\nonumber\\=\frac{\rho_l-\rho_g}{2\Sigma^2}
\left[\frac{1}{\theta(\theta-\alpha)^2}
-\frac{1}{\theta^3}\right]
\label{adsorption}
\end{eqnarray}
where $\rho_g$ and $\rho_l$ are the coexistence densities of the 
vapor and liquid phases, respectively. Close to the filling transition 
($\theta\to \alpha$), $\Gamma_w \sim 2(\rho_l-\rho_g)\langle l(0)\rangle_w^2/
\alpha$.

The roughness profile $\xi_\perp(x)$ (see Fig.\ref{fig1}) is defined as 
$\sqrt{\langle l^2(x)\rangle_w - \langle l(x)\rangle_w^2}$, where 
$\langle l^2(x)\rangle_w$ is given by:
\begin{eqnarray}
& &\langle l^2(x)\rangle_w = \frac{1}{2\Sigma^2 \theta^2}
\nonumber\\
&-&\Bigg(-\frac{1}{\Sigma(\theta-\alpha)}-\frac{1}{\Sigma\alpha}
+\frac{1}{\Sigma\theta}+\theta x\Bigg)\sqrt{\frac{x}{2\pi\Sigma}} 
\textrm{e}^{-\frac{\Sigma \theta^2}{2}x}\nonumber\\&+&
\left[\frac{\theta^2}{(\theta-\alpha)^2}-\frac{\theta^2}{\alpha^2}\right]
\frac{\textrm{e}^{2\Sigma \alpha x (\alpha - \theta)}}{4\Sigma^2\theta^2}
\textrm{erfc}\left(\sqrt{\frac{\Sigma x}{2}}
(\theta-2\alpha)\right)\nonumber\\
&-&\Bigg[\frac{1}{4\Sigma^2\theta^2}\left(-\frac{\theta^2}{(\theta-\alpha)^2}-
\frac{\theta^2}{\alpha^2}+2\right)-\frac{\theta x}{2\Sigma}\Bigg(\frac{\alpha}
{\theta-\alpha}\nonumber\\ &-&\frac{\theta-\alpha}{\alpha}\Bigg)
+\frac{x^2\theta^2}{2}\Bigg]
\textrm{erfc}\left(\sqrt{\frac{ \Sigma 
\theta^2}{2}x} \right)
\label{av2}
\end{eqnarray}
For general $n$, the following expression can be obtained by induction:
\begin{eqnarray}
\langle l^n(x)\rangle_w = \langle l^n\rangle_\pi \Bigg[1+ 
\frac{1}{2}\left(\frac{\theta^n}{(\theta-\alpha)^n}-\frac{\theta^n}{\alpha^n}
\right)\textrm{e}^{2\Sigma \alpha x (\alpha - \theta)}\nonumber\\
\times\textrm{erfc}\left(\sqrt{\frac{\Sigma x}{2}}
(\theta-2\alpha)\right)\Bigg]
+P_n(x)\sqrt{\frac{x}{2\pi\Sigma}} 
\textrm{e}^{-\frac{\Sigma \theta^2}{2}x}\nonumber\\
+Q_n(x)\textrm{erfc}\left(\sqrt{\frac{ \Sigma 
\theta^2}{2}x} \right)
\label{av3}
\end{eqnarray}
where $\langle l^n\rangle_\pi=n!/(2\Sigma \theta)^n$ and 
$P_n(x)$ and $Q_n(x)$ are polynomials in $x$ of order $n-1$ and $n$, 
respectively.

These expressions are only valid if $\theta > \alpha$ (for smaller values 
of $\theta$ the interface is unbound from the wedge). For $x\to 0$, 
Eq. (\ref{PDFx}) reduces to Eq. (\ref{PDF0}). On the other hand,
for $|x|\to \infty$, $P_w(l,x)$ decay to $P_\pi(l)\equiv 2 
\Sigma\theta \exp(-2 \Sigma \theta l)$. However, the scale over which 
this decay occurs depends on the value of $\alpha$. If 
$\theta \ge 2\alpha$, this scale is the planar correlation length 
$\xi_\parallel\equiv 2/\Sigma \theta^2$. 
However, if $\alpha < \theta < 2 \alpha$, the decay length 
is $\xi_F \equiv 1/2 \Sigma \alpha 
(\theta-\alpha)$ (our notation differs slightly from the one used in Ref.
\cite{Wood}). Note that $\xi_F$ is always larger than $\xi_\parallel$, and
diverges on approaching the filling transition. On the other hand, $\xi_F$ is
related geometrically with the wedge midpoint average interfacial height via
$\xi_F=\langle l(0)\rangle_w/\alpha \approx \langle l(0)\rangle_w/ \tan \alpha$
for small $\alpha$. 

It is amusing to note that Eq. (\ref{PDFx}) verifies the 
following differential relation:
\begin{eqnarray}
P_w(l,x)+\xi_F \left(\frac{\partial P_w(l,x)}{\partial x}\right)=L_\pi(l,x)
\nonumber\\ \equiv P_\pi(l)+\frac{1}{\theta}\frac{\partial}{\partial x}
\int_0^\infty dl_0 l_0 P_\pi (l_0, 0; l, x)
\label{fieldeq1}
\end{eqnarray}
where $L_\pi(l,x)$ is for contact binding potentials:
\begin{eqnarray}
L_\pi(l,x)=\Sigma\theta \textrm{e}^{-2\Sigma\theta l}\textrm{erfc}
\left(-\sqrt{\frac{\Sigma x}{2}}\theta+\sqrt{\frac{\Sigma}{2x}}l\right) 
\nonumber\\ +2\sqrt{\frac{\Sigma}{2\pi x}}
\textrm{e}^{-\left(\sqrt{\frac{\Sigma x}{2}}\theta+\sqrt{\frac{\Sigma}{2x}}l
\right)^2} 
\label{fieldeq2}
\end{eqnarray}
Note that the RHS of Eq. (\ref{fieldeq1}) depends only on the planar 
properties and, consequently, is independent of $\alpha$. 
It can be shown that Eq. (\ref{fieldeq1}) is obtained 
for \emph{any} binding potential if the LHS is expanded in powers of $\alpha$ 
and truncated at the lowest order term, which is independent of $\alpha$. 
Consequently, this differential field equation implies an infinite hierarchy 
of integro-differential relationships for the $2-$point planar correlation 
function. 
Alternatively Eq. (\ref{fieldeq1}) provides an elegant route to the 
calculation of  
any moment of the interfacial height. Multiplying Eq. (\ref{fieldeq1}) 
by arbitrary power of $l$ and integrating over all the possible values 
of $l$, the following differential equations are obtained:
\begin{eqnarray}
\langle l^n(x) \rangle_w + \xi_F \frac{d \langle l^n(x) \rangle_w}{dx}=
\int_0^\infty dl\ l^n L_\pi(l,x)\nonumber\\ \equiv
\langle l^n \rangle_\pi + \frac{1}{\theta} \frac{d}{dx}\langle l(0) l^n(x) 
\rangle_\pi
\label{moments}
\end{eqnarray}
where $\langle\ldots\rangle_w$ and $\langle\ldots\rangle_\pi$ mean
the average with the wedge and planar distribution function, respectively.
The RHS of Eq. (\ref{moments}) depends only in the planar distribution
functions, and consequently decays to $\langle l^n \rangle_\pi$ for
distances larger than $\xi_\parallel$. Close to the filling transition,
$\xi_F \gg \xi_\parallel$, and we can approximate 
Eq. (\ref{moments}) for $x\gtrsim \xi_F$ by:
\begin{equation}
\langle l^n(x) \rangle_w + \xi_F \frac{d \langle l^n(x) \rangle_w}{dx}
\approx \langle l^n \rangle_\pi 
\label{moments2}
\end{equation}
that has as a solution $\langle l^n(x) \rangle_w \approx 
\langle l^n \rangle_\pi + (\langle l^n(0)\rangle_w - \langle l^n \rangle_\pi)
\exp(-x/\xi_F)$. Taking into account that $\langle l^n(0)\rangle_w \gg
\langle l^n \rangle_\pi$ close to the filling transition, the approximate 
solution can be simplified even further to $\langle l^n(x) \rangle_w \approx
\langle l^n(0)\rangle_w\exp(-x/\xi_F)$ (which is equivalent to set
$\langle l^n \rangle_\pi = 0$ in Eq. (\ref{moments2}) ). These findings are
obviously in agreement with Eq. (\ref{av3}) and the asymptotic behavior 
of $P_w(l,x)$ for large $x$ and $\theta< 2 \alpha$ \cite{Wood}.

It is interesting to note that the moments obtained from the actual $1-$point
distribution function are the only solutions of Eq. (\ref{moments}) that: (a)
decay exponentially within a length scale $\xi_\parallel$ for $0<\alpha/\theta 
\ll 1$ and $x\to \infty$; (b) are analytical as a function of $\alpha$ for
$0\le \alpha < \theta$, in particular at the disorder point. 
The existence of the relationship Eq. (\ref{fieldeq1}), from which covariance
for the moments of the interfacial position profile at $x=0$ can be inferred
provided the (a) and (b) regularity conditions are fulfilled, leads us to 
speculate on the existence of a hidden symmetry of the hamiltonian 
that explains wedge covariance. However, the nature of such a symmetry 
(if any) is completely unknown. 

In the mean-field approximation, the average interfacial position profile 
for binding potentials characterized by a critical exponent $\alpha_s=0$ 
fulfills the following generalized covariance relationship \cite{Greenall}:
\begin{equation}
l(x)=l_\pi\left(\theta-\left|\frac{dl(x)}{dx}\right|\right)
\label{covprof}
\end{equation}
where $l(x)$ represents the (averaged) interfacial position at $x$, and
$l_\pi(\theta)$ is the planar (averaged) interfacial position for a given
contact angle $\theta$. Making the substitution $l(x)\to \langle l(x)
\rangle_w$, it is clear from Eq. (\ref{av1}) that this extended covariance is
not verified for $x\ne 0$ (even asymptotically when $x\to 0$ or $|x|\to 
\infty$). However, it is remarkable that there exists an analogous to 
Eq. (\ref{covprof}), given by Eq. (\ref{moments}) for $n=1$. 
\begin{figure}
\includegraphics[width=8.6cm]{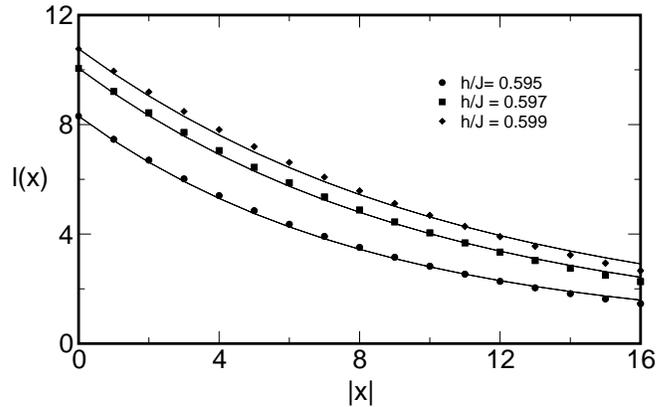}
\caption{Comparison between $\langle l(x)\rangle_w$ obtained in Ref.
\cite{Albano} by Ising model computer simulations for boundary magnetic
fields $h/J=0.595$ (circles), $h/J=0.597$ (squares) and $h/J=0.599$ 
(diamonds); and the approximation given by Eq. (\ref{compising}) 
(continuous lines). The Ising model parameters are: $\alpha=\pi/4$, the 
temperature $T=T_c/2$ and the bulk magnetic field $H_{bulk}=0$. The 
boundary magnetic field at the critical filling is $h_c/J=0.606$. The
lengths $|x|$ and $\langle l(x)\rangle_w$ are measured in lattice spacing
units. See text for explanation.\label{fig2}}
\end{figure}

To finish our discussion about the $1-$point distribution functions, we compare
our results with computer simulations of the 2D Ising model \cite{Albano}. 
Close to the filling transition point, we expect that the approximate 
solution to Eq. (\ref{moments2}) for $n=1$ will be generalized for arbitrary 
$\alpha$ to:
\begin{equation}
\langle l(x) \rangle_w \approx \frac{\langle l\rangle_\pi}
{\cos\alpha} + \left(\langle l(0) \rangle_w - \frac{\langle l\rangle_\pi}
{\cos\alpha}\right) \textrm{e}^{-x/\xi_F}
\label{compising}
\end{equation}
where now $\xi_F$ is defined as $\langle l(0) \rangle_w/\tan \alpha$. 
We have tested this approximation with the simulation results reported in Ref. 
\cite{Albano} (see Fig. \ref{fig2}). The symbols correspond to the simulation 
data obtained for an square $64\times 64$ Ising lattice with zero bulk magnetic
field and boundary magnetic fields $+h$ for the boundary rows ending at the 
lower left corner, and $-h$ for the remaining boundary rows. In this geometry,
$\alpha=\pi/4$. The temperature is set to $T=T_c/2$, where $T_c$ is the 
bulk critical temperature. For this temperature and $\alpha$ 
the critical filling transition occurs at $h_c/J=0.606$. Fig. \ref{fig2} shows
the computer simulation results for $h/J=0.595,0.597$ and $0.599$.
We have no direct estimation of $\langle l \rangle_\pi$. However, we have 
obtained $\langle l\rangle_\pi$ by fitting the simulation data with 
$|x|\le 16$ lattice spacings (in order to minimize the effect of the
upper left and lower right heterogeneous wedges) to the Eq. (\ref{compising}). 
The best fitting values are, in lattice spacing units, $\langle l
\rangle_\pi=0.314,0.335,0.436$ for $h/J=0.595,0.597,0.599$, respectively. 
As it can be seen, the fitting to the simulation data is quite good, despite
the crude approximations involved in Eq. (\ref{compising}).

Now we want to characterize the $2-$point correlations, in particular the 
correlations between the interfacial position above the wedge midpoint and the
corresponding to an arbitrary $x$, which are given by the following function:
\begin{eqnarray}
\langle (l(x)-\langle l(x)\rangle_w)(l(0)-\langle l(0)\rangle_w)\rangle_w
\equiv \nonumber\\
\langle l(x)l(0)\rangle_w-\langle l(x)\rangle_w\langle l(0)\rangle_w=
\frac{1}{2\Sigma} \left(\frac{\partial \langle l(x)\rangle_w}{\partial \alpha}
\right)
\label{corr1}
\end{eqnarray}
Substituting Eq. (\ref{av1}) into Eq. (\ref{corr1}), we obtain:
\begin{eqnarray}
\langle l(x)l(0)\rangle_w-\langle l(x)\rangle_w\langle l(0)\rangle_w
 = \sqrt{\frac{x}{2\pi\Sigma}}
\frac{(2\alpha-\theta)\textrm{e}^{-\frac{\Sigma \theta^2}{2}x}}
{2\Sigma\alpha (\theta-\alpha)}
\nonumber\\+
\frac{1}{8\Sigma^2\theta^2}\left(\frac{\theta^2}{(\theta-\alpha)^2}-
\frac{\theta^2}{\alpha^2}\right) 
\textrm{erfc}\left(\sqrt{\frac{ \Sigma \theta^2}{2}x}
\right)
\nonumber\\+\frac{\textrm{e}^{2\Sigma \alpha x (\alpha - \theta)}}
{8\Sigma^2\theta}
\Bigg[\frac{\theta}{(\theta-\alpha)^2}+\frac{\theta}{\alpha^2}\nonumber\\+
\frac{2\Sigma(\theta-2\alpha)^2\theta x}{\alpha(\theta-\alpha)}\Bigg]
\textrm{erfc}\left(\sqrt{\frac{ \Sigma x}{2}}
(\theta-2\alpha)\right)
\label{corr2}
\end{eqnarray}
This function decays exponentially to zero for large $x$. However, the 
characteristic correlation length $\xi_x$ (see Fig. \ref{fig1}) depends on 
$\alpha$: it is $\xi_\parallel$ for $\theta>2\alpha$ and $\xi_F$ if 
$\alpha<\theta<2\alpha$. Consequently, the disorder point not only 
introduce a new length scale for the average interfacial profile, but also 
for the interfacial fluctuations. 

\section{Results for the Kratzer binding potentials \label{sec4}}

The Kratzer binding potential (see Eq. (\ref{kratzer})) is the marginal 
case between the filling mean-field and filling fluctuation regimes. 
For such potentials the wedge midpoint probability distribution function 
also obeys wedge covariance Eq. (\ref{covariance}):
\begin{eqnarray}
P_w^1(l_0;\theta,\alpha)=\frac{[2\Sigma(\theta-\alpha)]^{2\phi+1}l_0^{2\phi}}
{\Gamma[2\phi+1]}\exp(-2\Sigma(\theta-\alpha)l_0)\nonumber\\
=P_\pi^1(l_0;\theta-\alpha)
\label{cov3}
\end{eqnarray}
It is possible to extend the transfer analysis and obtain exact results
for other quantities of interest. Consider, for example, the $1-$point 
probability distribution function $P_w(l,x)$. The Laplace transform 
$\tilde P_w(l;E)$ can be expressed as:
\begin{eqnarray}
\tilde P_w(l;E)&=&\int_0^\infty dl_0 \textrm{e}^{\Sigma
(2\alpha-\theta) l_0}\frac{(2\Sigma \theta )^{2\phi+1} 
(l_0 l)^\phi}{\Gamma[2\phi+1]}\nonumber\\
&\times& \exp(-\Sigma\theta l)\tilde{Z}_\pi(l_0,l,E-\Sigma \theta^2/2)
\label{laplacep1}
\end{eqnarray}
where $\tilde{Z}_\pi(l_0,l,E)$ is given by Eq. (\ref{laplacecoulomb}).
This reduces to:
\begin{eqnarray}
\tilde P_w(l;E)=\frac{l^\phi(2\Sigma \theta)^{2\phi+1}
\textrm{e}^{-\Sigma \theta l}}{\theta\Gamma[2\phi+1]\Gamma[2\phi]}
\kappa\Gamma\left[\phi\left(1-\kappa\right)\right]
\nonumber\\
\times \Bigg\{\int_0^\infty l_0^\phi \textrm{e}^{\Sigma(2\alpha-\theta)l_0}
W_{\kappa\phi,\phi-\frac{1}{2}}
\left(\frac{2\Sigma \theta l_0}{\kappa}\right)
\nonumber\\
M_{\kappa\phi,\phi-\frac{1}{2}}
\left(\frac{2\Sigma \theta l}{\kappa}\right)
-\int_0^l l_0^\phi \textrm{e}^{\Sigma(2\alpha-\theta)l_0}
\nonumber\\
\times \Bigg[W_{\kappa\phi,\phi-\frac{1}{2}}
\left(\frac{2\Sigma \theta l_0}{\kappa}\right) M_{\kappa\phi,\phi-\frac{1}{2}}
\left(\frac{2\Sigma \theta l}{\kappa}\right)
\nonumber\\
-W_{\kappa\phi,\phi-\frac{1}{2}}
\left(\frac{2\Sigma \theta l}{\kappa}\right) M_{\kappa\phi,\phi-\frac{1}{2}}
\left(\frac{2\Sigma \theta l_0}{\kappa}\right)\Bigg]\Bigg\}
\end{eqnarray}
where $\kappa\equiv 1/\sqrt{1-2E/\Sigma\theta^2}$.
The poles of $\tilde P_w(l;E)$ in the $E$ real positive semi-axis
are the characteristic inverse length
scales across the wedge of $P_w(l,x)$. Since the second integral is 
over a finite interval and the integrand does not diverges in that range, 
no new length scale emerges from it. For the first integral, we take into
account that \cite{book}:
\begin{eqnarray}
\int_0^\infty x^{\nu-1} \exp(-px)W_{\kappa,\mu}(ax)dx=
\nonumber\\
\frac{\Gamma[\mu+\nu+1/2]\Gamma[\nu-\mu+1/2]a^{\mu+\frac{1}{2}}}
{\Gamma[\nu-\kappa+1](p+a/2)^{\mu+\nu+\frac{1}{2}}}
\nonumber\\
\times _2F_1\left(\mu+\nu+\frac{1}{2},\mu-\kappa+\frac{1}{2};\nu-\kappa+1;
\frac{p-\frac{a}{2}}{p+\frac{a}{2}}\right)
\label{intwkm}
\end{eqnarray}
where $_2 F_1(a,b,c;x)$ is a hypergeometric function. If $\theta>2\alpha$, 
the integral does not introduce any new characteristic length . However, 
for $\alpha<\theta<2 \alpha$ a new singularity emerges for $\Sigma 
(\theta-2\alpha)+\Sigma\theta/\kappa=0$, i.e. $E=2\Sigma \alpha(\theta-
\alpha)=1/\xi_F$. Remarkably, $\xi_F$ has the same expression as for 
contact binding potentials, and is proportional (but not equal) to 
$\langle l(0)\rangle_w/\alpha$.

From this it follows that the non-thermodynamic singularity occuring at 
$\theta=2\alpha$ mentioned in the previous Section is not specific to 
contact potentials. A simple geometrical argument given in Ref.
\cite{Wood} explains why. The most relevant interfacial fluctuations 
are those where the interface leaves the substrate with a contact angle 
$\theta$ (relative to the tilted wall) at an arbitrary substrate point. 
If $\theta > 2 \alpha$, the other side of the wedge does not play any role 
and we can anticipate that the only length scale that controls the $1-$point 
distribution decay is $\xi_\parallel$. However, if $\theta < 2\alpha$, the 
interface will eventually reach the other substrate, and consequently we 
can expect the geometry to play an important role leading to the emergence 
of a new length scale. Formally, this non-thermodynamic singularity occurs 
when the following integrals that arise from the spectral expansion of 
$Z_\pi(l_1,l_2;x)$:
\begin{equation}
\int_0^\infty \psi_0(l) \exp(2\Sigma \alpha l) \psi^*_p(l) 
\label{disorder}
\end{equation}
become ill-defined. There, $\psi_p(l)$ are the scattering eigenstates 
with eigenvalues $E=p^2/2 \Sigma$ and $\psi_0(l)$ is the ground 
eigenstate. A straightforward WKB asymptotic analysis for the eigenfunctions
shows that, for $p\ne 0$, the integrals given by Eq. (\ref{disorder})
become ill-defined for $\theta<2\alpha$ for quite arbitrary choices of
binding potential. 

As $\theta/\alpha$ decreases, $\xi_F$ exceeds the intrinsic interfacial 
length scales $1/(E_i-E_0)$, and becomes the true correlation length 
across the wedge $\xi_x$ at an another disorder point when 
$\xi_F=\xi_\parallel$ (recall that $\xi_x=\xi_\parallel$ for $\theta/\alpha$ 
larger than the value at the disorder point). For the case of contact
binding potentials both non-thermodynamic singularities occur the same value
$\theta=2\alpha$. However, in general, the non-thermodynamic singularities
are distinct provided there are at least two bounded eigenstates of Eq. 
(\ref{schrodinger2}). For the pure Coulomb case ($\phi=1$) the second 
disorder point occurs at $\theta=4\alpha/3$.

Close to the new singularity $\xi_F^{-1}$ we found that: 
\begin{equation}
P_w(l;E)\sim \frac{1}{(\xi_F^{-1}-E)^{1+\frac{2\phi\alpha}{2\alpha-\theta}
}} \quad E\xi_F\to 1^-
\label{asymptoticpw}
\end{equation}
so $P_w(l,x)$ behaves asymptotically for large values of $x$ as 
$x^{\frac{2\phi\alpha}{2\alpha-\theta}}\exp(-x/\xi_F)$, provided that 
$\xi_\parallel < \xi_F$.

A field equation analogous to Eq. (\ref{fieldeq1}) can be found for
Kratzer potentials. Transfer matrix calculations for arbitrary binding 
potentials lead to the relation:
\begin{eqnarray}
\langle 0 | \textrm{e}^{2\Sigma\alpha l}|0\rangle\Bigg\{ 
\frac{\theta-\alpha}{\theta}
\left[P_w(l,x)+\xi_F \left(\frac{\partial P_w(l,x)}{\partial x}\right)
\right]\nonumber\\
-\int_0^{\infty}dl_0\frac{\bar{\psi}_0'(l_0)}{\Sigma\theta \bar{\psi}_0(l_0)}
P_w(l_0,0;l,x)\Bigg\}=L_\pi(l,x)
\nonumber\\ - \int_0^{\infty}dl_0\frac{\bar{\psi}_0'(l_0)}{\Sigma\theta 
\bar{\psi}_0(l_0)}P_\pi(l_0,0;l,x)
\label{fieldeq-ex}
\end{eqnarray}
where $\bar{\psi}_0(l_0)\equiv \psi_0(l_0)\exp(\Sigma \theta l_0)$ and
$\bar{\psi}_0'(l_0)$ is its derivative respect to $l_0$ (recall that
$\psi_0(l_0)$ is the ground state eigenfunction). For Kratzer potentials,
$\bar{\psi}_0(l_0)\propto l_0^\phi$, so Eq. (\ref{fieldeq-ex}) can be 
expressed as:
\begin{eqnarray}
\left(\frac{\theta}{\theta-\alpha}\right)^\phi \frac{\partial P_w(l,x)}
{\partial \alpha}\nonumber\\+\frac{\partial}{\partial \alpha}\Bigg[\xi_F 
\left(\frac{\theta}{\theta-\alpha}\right)^\phi \frac{\partial P_w(l,x)}
{\partial x}\Bigg]=0
\label{fieldeq-ex2}
\end{eqnarray}
As for the contact binding potential case, some interesting quantities can 
be evaluated from this expression. For example, the wedge adsorption is 
found to be:
\begin{equation}
\Gamma_w=(2\phi+1)(\phi+1)\Gamma^{CP}
\label{adsorption2}
\end{equation}
where $\Gamma^{CP}$ is the adsorption corresponding to the contact binding 
potential Eq. (\ref{adsorption}).

\section{The breather mode picture\label{sec5}}

In order to understand the origin of the new correlation length $\xi_F$ 
we identified in previous Sections, we recall the definition of the 
$2-$point distribution function for $x_2>x_1\ge 0$, Eq. (\ref{wedgecorr2}). 
This expression can be written in the following way:
\begin{equation}
P_w^c(l_2,x_2|l_1,x_1)=P_\pi^c(l_2,x_2|l_1,x_1)\equiv 
P_\pi^c(l_2,x_2-x_1|l_1,0)
\label{conditional}
\end{equation}
where $P_{w}^c(l_2,x_2|l_1,x_1)$ and $P_{\pi}^c(l_2,x_2|l_1,x_1)$ are, 
respectively, the wedge and planar conditional probability of the 
interface being at a relative height $l_2$ from the substrate at $x_2$, 
provided that the interface is pinned at a relative height $l_1$ at $x_1$, 
defined as:
\begin{equation}
P_i^c(l_2,x_2|l_1,x_1)= \frac{P_i(l_1,x_1;l_2,x_2)}
{P_i(l_1,x_1)}
\label{defconditional}
\end{equation}
where the subscript $i$ indicates if this probability is considered in 
the wedge ($i=w$) or in the planar ($i=\pi$) geometry.

\begin{figure}
\includegraphics[width=8.6cm]{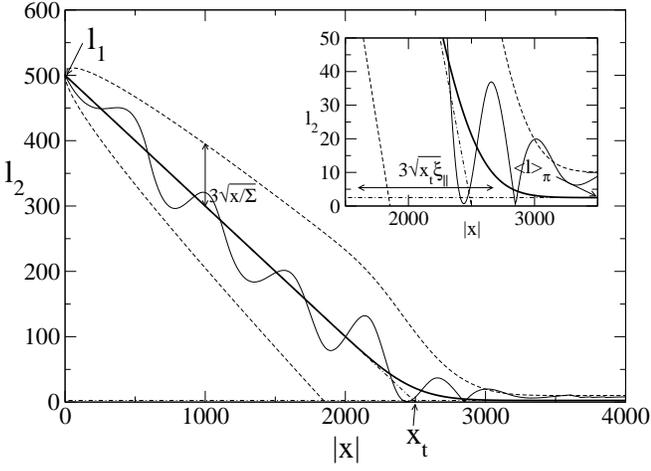}
\caption{Illustration of a typical interfacial configuration pinned at $l_1
\gg \langle l \rangle_\pi$ for $x=0$ (thin continuous line). We have set
$\Sigma=1$ (it defines the length scale), $\theta=0.2$ and $l_1=500$. The 
thick continuous line corresponds to the conditional average profile $\langle 
l_2\rangle^c(l_1,x)$, and the dotted lines correspond to $\max(0,\langle 
l_2\rangle^c(l_1,x)\pm 3 \xi_\perp^c(l_1,x))$ where $\xi_\perp^c(l_1,x)$ is
the conditional roughness. Any interfacial configuration has a probability 
of at least 95\% of being between the dotted lines. Inset: an enlargement of 
the area around $x_t=l_1/\theta$. Other characteristic length scales are
represented. See text for explanation.\label{fig3}}
\end{figure}

In view of the identity between the wedge and planar conditional 
probability distribution functions we first consider the of a planar substrate.
The conditional probability can be obtained as:
\begin{equation}
P^c_\pi(l_2,x|l_1,0)=\frac{\psi_0^*(l_2)}{\psi_0^*(l_1)}\textrm{e}^{-\frac{
\Sigma \theta^2}{2}x}Z_\pi(l_1,l_2;x)
\label{defconditional2}
\end{equation}
For contact binding potentials, Eq. (\ref{defconditional2}) can be written 
explicitely as:
\begin{eqnarray}
P^c_\pi(l_2,x|l_1,0)=\sqrt{\frac{\Sigma}{2\pi x}}\textrm{e}^{-
\frac{\Sigma(l_2-l_1+\theta x)^2}{2x}}\nonumber\\
+\textrm{e}^{- 2\Sigma\theta l_2}\Bigg[\sqrt{\frac{\Sigma}{2\pi x}}
\textrm{e}^{-\frac{\Sigma(l_1+l_2-\theta x)^2}{2x}}
\nonumber\\
+ \Sigma\theta \textrm{erfc}\left(\sqrt{\frac{ \Sigma}{2\pi x}}
(l_1+l_2-\theta x)\right)\Bigg]
\label{defconditional3}
\end{eqnarray}
If $l_1$ is very large compared with $\langle l\rangle_\pi\equiv
1/2\Sigma \theta$, we can identify two different behaviors of 
$P^c_\pi(l_2,x|l_1,0)$ as a function of $l_2$ (see Fig. \ref{fig3}). 
If $x<l_1/\theta$, the conditional probability is
basically the free interface conditional probability that fluctuates
around an average value $\langle l_2(x)\rangle=l_1-\theta x$, with a 
standard deviation of the order of $\sqrt{x/\Sigma}$. For 
$x>l_1/\theta$, the conditional probability becomes the $1-$point planar 
planar distribution function $P_\pi(l_2)=2\Sigma\theta\exp(-2\Sigma\theta 
l_2)$, completely uncorrelated to the value of $l_1$. The transition 
between the two regimes occur in an $x$ interval around $x_t=l_1/\theta$ 
which has a width of order of $\sqrt{2l_1/\Sigma\theta^3}\equiv\sqrt{x_t 
\xi_\parallel}$. These results are confirmed by the exact evaluation of the
first moments of the conditional probability:
\begin{equation}
\langle l_2^n \rangle^c(l_1,x)=\int_0^\infty dl_2\ l_2^n P^c_\pi(l_2,x|l_1 0)
\label{momentcond}
\end{equation}
The average conditional interfacial profile, which corresponds to $n=1$, 
is given by:
\begin{eqnarray}
\langle l_2 \rangle^c(l_1,x)=(l_1-\theta x)+\sqrt{\frac{\Sigma}{2\pi x}}
\textrm{e}^{-\frac{\Sigma(l_1-\theta x)^2}{2x}}\nonumber\\
+ \left[\frac{1}{4\Sigma \theta}-\frac{l_1-x\theta}{2}\right]
\textrm{erfc}\left(\sqrt{\frac{ \Sigma}{2\pi x}}(l_1-\theta x)\right)
\nonumber\\-\frac{\textrm{e}^{2\Sigma\theta l_1}}{4\Sigma \theta}
\textrm{erfc}\left(\sqrt{\frac{ \Sigma}{2\pi x}}(l_1+\theta x)\right)
\label{momentcond2}
\end{eqnarray}
and the conditional roughness $\xi_\perp^c(l_1,x)$ is defined as $\sqrt{
\langle l_2^2 \rangle^c - \langle (l_2 \rangle^c)^2}$, where $\langle l_2^2 
\rangle^c(l_1,x)$ can be written as:
\begin{eqnarray}
\langle l_2^2 \rangle^c(l_1,x)=\left[(l_1-\theta x)^2+\frac{x}{\Sigma}\right]
-\Bigg(\frac{1}{\Sigma\theta}-l_1+\nonumber\\
\theta x\Bigg)\sqrt{\frac{\Sigma}{2\pi x}}\textrm{e}^{-\frac{\Sigma(l_1-
\theta x)^2}{2x}}+ \Bigg[\frac{x}{2\Sigma} -\frac{1}{4\Sigma^2 \theta^2}+
\nonumber\\ \frac{(l_1-x\theta)^2}{2}\Bigg]
\textrm{erfc}\left(\sqrt{\frac{ \Sigma}{2\pi x}}(l_1-\theta x)\right)
\nonumber\\+\left[x\theta+l_1-\frac{1}
{2\Sigma\theta}\right]\frac{\textrm{e}^{2\Sigma\theta l_1}}{2\Sigma\theta}
\textrm{erfc}\left(\sqrt{\frac{ \Sigma}{2\pi x}}(l_1+\theta x)\right)
\label{momentcond3}
\end{eqnarray}
We obtain two main conclusions from these results when $l_1\gg \langle l
\rangle_\pi$. First, the interfacial positions are highly correlated to the 
the central one for $|x|<l_1/\theta$. Secondly, the intrinsic interfacial
fluctuations are small in this $x$ range compared to the conditional average
value. Actually, if we set $l_1$ as the length scale, the rescaled conditional
probability distribution function $\tilde{P}_\pi^c(l_2/l_1,x/l_1|1,0)\equiv
l_1 P_\pi^c(l_2,x|l_1,0)$ behaves as:
\begin{eqnarray}
\tilde{P}_\pi^c(l_2/l_1,x/l_1|1,0)\to 
\delta\left(\frac{l_2-l_1+\theta x}{l_1}\right) H(l_1-\theta x)
\nonumber\\
+\delta\left(\frac{l_2}{l_1}\right) H(\theta x-l_1)
\label{asympt1}
\end{eqnarray}
when $\Sigma \theta l_1 \to \infty$. We expect this result to be valid for 
any potential and also for random bond disorder, since in all these cases 
the wandering exponent for the free interface $\zeta<1$. 
this can be checked for the marginal
$1/l$ potential. The Laplace transform of the conditional
probability distribution as:
\begin{eqnarray}
{\cal L}\left[P_\pi^c(l_2,x|l_1,0)\right]
\equiv \int_0^\infty dx\ \textrm{e}^{Ex}
P_\pi^c(l_2,x|l_1,0)\nonumber\\
=\frac{\psi_0^*(l_2)}{\psi_0^*(l_1)}\tilde{Z}_\pi(l_1,l_2,
E-\Sigma\theta^2/2)
\label{coulombcond}
\end{eqnarray}
For $\Sigma \to \infty$ at \emph{fixed} $E, \theta, l_1$ and $l_2$ , 
and taking into account Eq. (\ref{laplacecoulomb}) and that the ground 
state eigenfunction $\psi_0(l)\propto l^\phi \exp(-\Sigma\theta l)$, 
we obtain the following behavior for the Laplace transform of the 
conditional probability distribution function:
\begin{equation}
{\cal L}\left[P_\pi^c(l_2,x|l_1,0)\right]\to \frac{1}{\theta}
H(l_1-l_2)\textrm{e}^{E(l_1-l_2)/\theta}-\frac{\delta(l_2)}{E}
\textrm{e}^{El_1/\theta}
\label{asympt2}
\end{equation} 
The Laplace transform can be inverted, leading to Eq. (\ref{asympt1}).

To proceed, we return to our discussion about the wedge geometry. 
Due to the presence of the boost factor $\exp(2\Sigma\alpha l)$ in
the midpoint probability distribution function, the midpoint
interfacial height is almost always further from the substrate 
than the mean wetting layer thickness $\langle l \rangle_\pi$ for \emph{any} 
binding potential.
If we assume that the conditional probability distribution function is given
by Eq. (\ref{asympt1}), which corresponds to as neglecting the intrinsic 
interfacial
fluctuations around the conditional interfacial profile, we can capture
the main features of both the average interfacial profile and the correlations
along the wedge for contact binding potentials. Actually, this picture is 
completely equivalent to the 2D wedge breather mode 
model \cite{Parry4,Parry2}. 

The average interfacial profile can be written as:
\begin{eqnarray}
\langle l(x)\rangle_w= \int_0^\infty dl_1 P_w(l_1,0)\left[\int_0^\infty
dl_2 l_2 P_\pi^c (l_2,x|l_1,0)\right]\nonumber\\
\approx
\int_{\theta x}^\infty dl_1 P_w(l_1,0)(l_1-\theta x)=
\int_0^\infty s P_w(s+\theta x,0) ds
\label{breatherlx}
\end{eqnarray}
The behavior of $\langle l(x)\rangle_w$ for large $x$ is dominated by the
large $l$ asymptotics of $P_w(l,0)$. The latter can be obtained by
taking into account Eq. (\ref{wedgecorr3}) for $m=n=0$ and 
making use of the WKB approximation for the $1-$point planar distribution 
function:
\begin{eqnarray}
P_\pi(l)\sim \frac{1}{\sqrt{1+\frac{2W(l)}{\Sigma\theta^2}}}
\exp\left(-2\Sigma \theta \int^l dt\sqrt{
1+\frac{2 W(t)}{\Sigma \theta^2}}\right)
\nonumber\\
\sim \textrm{e}^{-2\Sigma \theta l}\exp\left(-2\int^l dt \frac{W(t)}{\theta}
\right)
\quad l\to \infty
\label{WKB}
\end{eqnarray}
The first thing we can see is that, for large $x$, the decay of $\langle l 
(x)\rangle_w$ in this approximation is controlled by an exponential term 
$\exp[-2\Sigma\theta(\theta-\alpha)x]$. So, a new length scale $\xi_F^*$ is 
defined as $1/2\Sigma\theta(\theta-\alpha)$. Close to the filling transition, 
$\xi_F^*=\xi_F-1/2\Sigma\alpha\theta\sim \xi_F + {\cal O}(1)$.

Depending on the large $l$ behavior of the (attractive) binding
potentials, different situations can arise \cite{Parry3}. The filling mean
field regime is characterized by binding potentials that decay to zero
as $1/l^p$ where $p<1/\zeta-1$, implying $\zeta<1$ for thermal disorder 
(the wandering exponent $\zeta=1/2$). A saddle point calculation
shows that close to the filling transition $\langle l(0) \rangle_w 
\sim 1/\Sigma (\theta-\alpha)^p$. As $\theta \to \alpha$, the relevant
length scale in the $x$ direction, $\langle l(0) \rangle_w/\theta \gg
\xi_F^*$, so the latter length scale is irrelevant (in fact, intrinsic 
interfacial fluctuations that we neglected can be more important). 

For $p=1$, both length scales become of the same order, and consequently 
$\langle l(x)\rangle_w \sim \langle l(0)\rangle_w f(x/\xi_F^*)
\exp(-x/\xi_F^*)$, where $f(x)$ diverges at most algebraically, and 
depends on the detailed structure of the binding potential through 
the short distance $l$ dependence of $P_w(l,0)$. For a pure $1/l$ potential, 
$f(x)=(1+2x/3+x^2/6)$. This expression verifies the differential equation
for $\langle l(x)\rangle_w$ that arises from Eq. (\ref{fieldeq-ex2}) in 
the scaling limit.

The filling fluctuation regime corresponds to potentials with $p>1$, 
and is characterized by universal critical exponents and scaling
functions. Indeed in the critical regime the scaling behavior is the same as 
that found for contact binding potentials. For $x \to \infty$, we find 
that asymptotically $\langle l(x) \rangle_w \sim \langle l(0) \rangle_w 
\exp(-x/\xi_F^*)$. 
This solution agrees with the asymptotics of $\langle l(x)\rangle_w$ for 
contact binding potentials when $\theta \to \alpha$, although with a decay 
length slightly smaller. However, the behaviour is asymptotically correct
if we assume that $\xi_F^*\equiv \xi_F$.

For the correlation functions, we have:
\begin{equation}
\langle l(x)l(0)\rangle_w -\langle l(x)\rangle_w \langle l(0)\rangle_w=
\int_0^\infty dl_1 l_1 P_w(l_1,0)\Delta(l_1,x) 
\label{corrbreather}
\end{equation}
where $\Delta(l_1,x)$ is defined as:
\begin{equation}
\Delta (l_1,x) = \int_0^\infty dl_2 l_2 \left[P_\pi^c(l_2,x|l_1,0)-
P_w(l_2,x)\right]
\label{defdelta}
\end{equation}
In the breather mode approximation, $\Delta (l_1,x)$ can be obtained as:
\begin{equation}
\Delta (l_1,x) \approx (l_1-\theta x) H(l_1-\theta x) - \langle l(x) \rangle_w
\label{deltabreather}
\end{equation}
We find different behaviors depending on the value of $p$. In the 
filling mean field regime, $\Delta (l_1,x)$ is negligible in this scale. 
For the filling fluctuation regime, the correlation function decays as:
\begin{equation}
\langle l(x)l(0)\rangle_w -\langle l(x)\rangle_w \langle l(0)\rangle_w
\sim \langle l(0)\rangle_w^2\left(1+\frac{x}{\xi_F^*}\right)\textrm{e}^
{-\frac{x}{\xi_F^*}}
\label{flucbreather}
\end{equation}
and again is in agreement with the behavior of the exact correlation function
for contact binding potentials Eq. (\ref{corr2}) when $x\to \infty$ and 
$\theta \to \alpha$ (assuming again that $\xi_F^*\equiv \xi_F$). 
Finally, for the marginal case $p=1$ the behavior of the correlation 
function is predicted to be for $x\to \infty$ as $\langle l(0)
\rangle_w^2 g(x/\xi_F^*) \exp(-x/\xi_F^*)$, where $g(x)$ is a function 
that diverges at most algebraically.

Another quantity of interest is the midpoint local 
susceptibility $\chi_w(l)$, defined as:
\begin{eqnarray}
\chi_w(l)=\frac{\partial \rho(l)}{\partial h}\Bigg |_{h=0}
=2(\rho_l-\rho_v)
\int_l^\infty ds P_w(s,0)\bar{\Delta}(s)
\label{suscept}
\end{eqnarray}
where $\bar{\Delta}(l)\equiv \int_0^\infty dx \Delta(l,x)$.
In the breather mode approximation and in the filling fluctuation regime,
$\bar{\Delta}(s)$ has the following expression:
\begin{equation}
\bar{\Delta}(l)=\frac{1}{\theta}\left(\frac{l^2}{2}-\langle l(0)\rangle_w^2
\right)
\label{deltabar}
\end{equation}
which is \emph{exact} for contact binding potentials. This expression, 
together with the midpoint wedge covariance Eq. (\ref{covariance}), leads
to the covariance relationship between the local susceptibilities 
\cite{Parry3}:
\begin{equation}
\chi_w(l;\theta,\alpha)=\frac{\theta-\alpha}{\theta}\chi_\pi(l,\theta-\alpha)
\label{susceptcov}
\end{equation}
where $\chi_\pi(l,\theta)$ is the local susceptibility corresponding to the
planar geometry for a contact angle $\theta$. 

Finally, we note that the breather mode picture has direct consequences
for the scaling of the interfacial profile in the filling fluctuation regime. 
To see this, recall that the wedge midpoint probability distribution 
function scales as \cite{Parry3}:
\begin{equation}
P_w(l)=\frac{1}{\langle l(0)\rangle_w}\Lambda\left(\frac{l}{\langle l(0)
\rangle_w} \right)
\label{scaling1}
\end{equation}
where $\Lambda(s)$ is a universal function and, due to covariance, is the
same as the scaling function for the corresponding planar $1-$point probability
distribution function. Complementing the scaling of the probability
distribution function is the position dependence of the interfacial 
profile, which we anticipate satisfies:
\begin{equation}
\langle l(x)\rangle_w=\langle l(0) \rangle_w \phi\left(\frac{\theta x}
{\langle l(0)\rangle_w}\right)
\label{scaling2}
\end{equation}
where $\phi(s)$ is another universal function.
In the breather mode picture, the interface is infinitely stiff in the
filled region implying that the scaling functions 
$\Lambda(s)$ and $\phi(s)$ are related via:
\begin{equation}
\phi(s)=\int_s^\infty d\bar s \Lambda(\bar s) (\bar s-s)
\label{scaling3}
\end{equation}
or equivalently:
\begin{equation}
\phi''(s)=\Lambda(s) 
\label{scaling4}
\end{equation}
A remarkable consequence of this relation is that the behavior of the 
interfacial profile close to the midpoint is determined by the short distance
behavior of the wedge midpoint $1-$point probability distribution function. 
Since $\Lambda(s)\sim s^{\frac{1}{\zeta}-2}$ as $s\to 0$ \cite{Parry3}, we have 
$\phi(s)\sim 1 - |s| + A|s|^{\frac{1}{\zeta}}$ for small $s$. Note that
the first two terms are needed to preserve the continuity of the
true interfacial profile $\langle y(x)\rangle_w$ and its derivative 
at the wedge midpoint. 
This result suggests that the interface behaves, for small values of $x$, 
as a random walk of $x$ as a function of $z$. This prediction is consistent
with the behaviour of $\langle l(x)\rangle_w$ for contact binding potentials
in the scaling limit $\theta\to \alpha$, $\xi_\parallel/\xi_F\to 0$ but
$x/\xi_F$ finite.  

Finally, note that Eq. (\ref{scaling4}) is also obeyed by the (non-universal) 
scaling functions corresponding to the marginal case. 

\section{Renormalization group approach to the critical filling transition
\label{sec6}} 

In this Section we will justify the critical properties of the filling 
transition using a renormalization group framework. Specifically we 
will generalize an exact decimation functional renormalization group 
procedure previously used to study 2D critical 
wetting \cite{Huse,Lipowsky,Spohn}. Our transfer matrix 
results show that geometry plays a fundamental role in determining the 
critical behavior, so we anticipate that the appropiate renormalization 
group procedure must preserve the wedge shape. This implies that the 
effective wandering exponent $\zeta$ determining the rescaling of the
interfacial height $l$ must be $\zeta=1$. This contrasts with the value 
$\zeta=1/2$, which is appropiate for free interfaces and also planar wetting
transtions. We will see that this choice leads naturally 
to the breather mode picture of the filling transition, implying that 
interfacial fluctuations are irrelevant except for those that determine 
the wedge midpoint interfacial position.

Before introducing the renormalization group scheme, we generalize some of the 
results of previous Sections. The set of $(2n+1)-$point distribution 
functions that 
includes the midpoint interfacial position can be obtained in terms of the 
planar case counterpart by Eq. (\ref{wedgecorr3}). If we set $\theta=\alpha$, 
it is clear from that expression that $P_w(-n;\ldots;n)\equiv 0$ at the 
critical filling transition for \emph{any} value of $n$. However, all the 
correlation functions decay at the same rate, since Eq. (\ref{wedgecorr3}) 
can be rewritten as:
\begin{equation}
P_w(-n;\ldots;n)=P_w(0)\frac{P_\pi(-n;\ldots;n)}{P_\pi(0)}
\label{npointcorr1}
\end{equation}
Consequently, the \emph{conditional} $(2n+1)-$point probability 
distribution function remains finite at the filling transition. The only 
relevant operator (in a renormalization group sense) should be related only 
to the $1-$point probability distribution function at the midpoint.
Taking into account Eq. (\ref{corrplanar}) and the definition of the 
$2-$point conditional probability distribution function 
Eq. (\ref{defconditional2}), we can rewrite Eq. (\ref{npointcorr1}) as
\begin{eqnarray}
P_w(-n;\ldots;n)&=&P_w(0)\prod_{i=0}^{n-1}P_\pi^c(l_{i+1},x_{i+1}-x_i|
l_i,0)\nonumber\\ &\times& \prod_{i=-n}^{-1} P_\pi^c(l_{i},x_{i+1}-
x_i|l_{i+1},0)
\label{npointcorr2}
\end{eqnarray}
where we have chosen the ground state eigenfunction to be real and positive.
The $2-$point conditional probability distribution function has a non-trivial 
limit when $\Sigma \theta l_1\to \infty$ (see Eq. (\ref{asympt1})).
Our goal will be to find a renormalization group scheme in which the 
$2-$point conditional probability distribution function converges to this limit,
and the only relevant operator is related to the wedge midpoint $1-$point
distribution function.

Let us consider a discrete version of the interfacial hamiltonian Eq. 
(\ref{hameff}):
\begin{equation}
\beta H = \sum_{i=-n}^{n-1} \left\{ \frac{\Sigma}{2}(z_{i+1}-z_i)^2+
W(z_i- i \alpha)\right\}
\label{discretehameff}
\end{equation}
where the spacing between sites $a=1$ defines the length unit for $l$, 
$\Sigma^{-1}$, etc. Using a similar transformation to the continuous case, Eq. 
(\ref{discretehameff}) can be written as:
\begin{eqnarray}
\beta H[l]=
2n \frac{\Sigma \alpha^2}{2}+ 2\Sigma \alpha l_n- 2\Sigma \alpha l_0
\nonumber\\ + \sum_{i=-n}^{n-1}\left\{\frac{\Sigma}{2}(l_{i+1}-l_i)^2+ 
\frac{W(l_i)+W(l_{i+1})}{2}\right\}
\label{discretehameff2}
\end{eqnarray}
where periodic boundary conditions have been applied ($l_n=l_{-n}$).
To simplify our discussion, we will consider $n\to \infty$ and neglect the
boundary effects. The probability of any interfacial configuration is given by:
\begin{equation}
P_w(\{l_i\})=\frac{\textrm{e}^{-\beta H}}{Z}=
\textrm{e}^{2\Sigma \alpha l_0+\beta f_W}\prod_{i=-\infty}^{\infty} 
\textrm{e}^{-\beta {\cal \tilde H}(l_i,l_{i+1})}
\label{probdisc}
\end{equation}
where ${\cal \tilde H}(l_i,l_{i+1})$ is defined as:
\begin{eqnarray}
\beta {\cal \tilde H}(l_i,l_{i+1})=\beta f_s+ \frac{\Sigma}{2}(l_{i+1}-l_i)^2+ 
\frac{W(l_i)+W(l_{i+1})}{2} 
\label{defcalh}
\end{eqnarray}
where $f_s$ is the planar surface free energy per unit length and is 
related to the contact angle corresponding to the binding potential $W(l)$
via:
\begin{equation}
\beta f_s = \frac{1}{2}\ln\left(\frac{\Sigma}{2\pi}\right)-\frac{\Sigma 
\theta^2}{2}
\label{fs}
\end{equation}
For $\theta=0$, $\beta f_s$ converges towards the free interface free energy
per unit length. Note that in the continuum limit Eq. (\ref{propagator}), this
term is absorbed in the path measure. On the other hand, $\beta f_W$ is 
defined as:
\begin{equation}
\beta f_W=\lim_{n\to \infty} \left\{\left(-{\ln Z}+2n\frac
{\Sigma \alpha^2}{2}\right) - 2n\beta f_s \right\}
\label{deffw}
\end{equation}
that corresponds to the wedge excess free energy (we suppose that this quantity
is well-defined). Note that ${\cal \tilde H}(l_i,l_{i+1})$ is invariant
under an exchange of its arguments.

Let us now consider a decimation procedure, similar to the one used for
the study of 2D critical wetting. We group the sites in blocks of
$b$ units, keeping the first one and integrating over all the interfacial 
positions of the remaining $b-1$ sites in the block. Since the site $i=0$ 
plays a special role, the sites to be kept in each decimation step 
are those with $i=jb$, with $j\in \mathbb{Z}$. After that, we rescale $x$ 
positions by a factor $b$ and $l$ by a factor $b^\zeta$, i.e.:
\begin{equation}
x \to x'=\frac{x}{b} \quad;\quad
l_i \to l_j'=\frac{l_{jb}}{b^\zeta}
\label{RGtrans}
\end{equation}
The new hamiltonian $\beta {\cal \tilde H'}(l_j',l_{j+1}')$ is defined as:
\begin{eqnarray}
\textrm{e}^{-\beta {\cal \tilde H'}(l_j',l_{j+1}')}=
b^\zeta \int_0^{\infty}dl_1 
\textrm{e}^{-\beta{\cal \tilde H}(b^\zeta l_j',l_1)}
\nonumber\\ \ldots \int_0^\infty dl_{b-1}
\textrm{e}^{-\beta{\cal \tilde H}(l_{b-1},b^\zeta l_{j+1}')}
\label{RGtrans2}
\end{eqnarray}
Note that the renormalized hamiltonian $\beta {\cal \tilde H}'(l_1,l_2)$ is 
symmetrical under an exchange of its arguments provided that the
original $\beta {\cal \tilde H}(l_1,l_2)$ is also symmetrical (even if
it is not defined as Eq. (\ref{defcalh})). This procedure is iterated, leading
to a sequence of renormalized hamiltonians. 

In order to complete the description of the renormalization group (RG) 
procedure we should give the transformation rules for $\alpha$ and 
$\beta f_W$. First we revisit the planar geometry ($\alpha=f_W=0$). 
We will consider the value of the exponent $\zeta$ arbitrary, unlike in 
Refs. \cite{Huse,Lipowsky,Spohn}, where $\zeta=1/2$. In general, after an 
arbitrary number of RG steps, we can write any 
$\beta {\cal \tilde H}(l_i,l_{i+1})$ as:
\begin{eqnarray}
\beta {\cal \tilde H}(l_i,l_{i+1})=\frac{\Sigma}{2}(l_{i+1}-l_i)^2+ 
\tilde{W}(l_i,l_{i+1})\nonumber\\
+ \frac{1}{2}\ln \left(\frac{\Sigma}{2\pi}\right)-\frac{\Sigma \theta^2}{2} 
\label{defcalh2}
\end{eqnarray}
where $\tilde{W}(l_i,l_{i+1})$ is a symmetric function under exchange of
its arguments. Obviously, in principle this function need not decay to zero
when both arguments are large (as it does in Eq. (\ref{defcalh})). 
However let us suppose that it decays as $-A/[(l_i+l_{i+1})/2]^p$ when
$l_i,l_{i+1}\to \infty$. After making a RG step, we would like to find the 
asymptotic behavior of the renormalized $\tilde{W}'(l_j',l_{j+1}')$ for 
large enough $l_j'$, $l_{j+1}'$. Expanding the RHS of Eq. (\ref{RGtrans2})
and keeping terms up to first order in $W$ (since $l_j'$ and $l_{j+1}'$ are
large, the values of $l_1\ldots l_{b-1}$ that contribute most to the integral
are also very large), we find that:
\begin{eqnarray}
\textrm{e}^{-\beta {\cal \tilde H'}}\approx
b^{\zeta} \textrm{e}^{b\frac{\Sigma \theta^2}{2}}
\int_0^\infty dl_1 \sqrt{\frac{\Sigma}{2\pi}}
\exp\left(-\frac{\Sigma (b^{\zeta} l_j'-l_1)^2}{2}\right)\nonumber\\\ldots
\int_0^\infty dl_{b-1} \sqrt{\frac{\Sigma}{2\pi}}
\exp\left(-\frac{\Sigma (l_{b-1}- b^{\zeta} l_{j+1}')^2}{2}\right)
\nonumber\\ \times \Bigg[1-W(b^\zeta l_j',l_1)-
W(l_{b-1},b^\zeta l_{j+1}')\nonumber\\
-\sum_{i=1}^{b-2} W(l_i,l_{i+1})\Bigg]
\label{largelasymp}
\end{eqnarray}
The lowest order term (corresponding to set $W=0$) can be estimated for 
large $l_j',l_1,\ldots,l_{b-1},l_{j+1}'$ by extending the lower integration
limits to $-\infty$, and has the value:
\begin{eqnarray}
b^{\zeta} \textrm{e}^{b\frac{\Sigma \theta^2}{2}}
\int_{-\infty}^\infty dl_1 \sqrt{\frac{\Sigma}{2\pi}}
\exp\left(-\frac{\Sigma (b^{\zeta} l_j'-l_1)^2}{2}\right)\nonumber\\\ldots
\int_{-\infty}^\infty dl_{b-1} \sqrt{\frac{\Sigma}{2\pi}}
\exp\left(-\frac{\Sigma (l_{b-1}- b^{\zeta} l_{j+1}')^2}{2}\right)
\nonumber\\
=\textrm{e}^{b\frac{\Sigma \theta^2}{2}} \sqrt{\frac{\Sigma b^{2\zeta -1}}
{2\pi}} \exp\left(-\frac{\Sigma b^{2\zeta -1}(l_j'-l_{j+1}')^2}{2}\right)  
\label{largeasymp2}
\end{eqnarray}
So, in order that the renormalized binding potential $W'$ decays to zero
at large values of both its arguments, the interfacial stiffness and the
contact angle must transform as:
\begin{equation}
\Sigma \to \Sigma'=\Sigma b^{2\zeta-1} \quad;\quad
\theta \to \theta'=\theta b^{1-\zeta}
\label{RGtrans3}
\end{equation}
Two comments are pertinent at this point. Firstly, the transformation of the 
interfacial stiffness is also valid for a free interface. Secondly, the change
in the contact angle has a geometrical interpretation, since its scaling
Eq. (\ref{RGtrans3}) corresponds exactly to the change of small angles under 
the coordinates scaling Eq. (\ref{RGtrans}). Thus it is reasonable to expect
that $\alpha$ must change in the same way.

If we take into account the first order in $W$, we can characterize the 
decay of the renormalized potential at large values of $l_j'$ and $l_{j+1}'$.
For simplicity, we consider the case $b=2$. The renormalized binding potential
has the following expression:
\begin{eqnarray}
W'(l_j',l_{j+1}')\approx \sqrt{\frac{\Sigma}{\pi}} \int_{-\infty}^0 dl_1
\textrm{e}^{-\Sigma\left(l_1-2^\zeta\frac{l_j+l_{j+1}}{2}\right)^2} \nonumber\\
+\sqrt{\frac{\Sigma}{\pi}} \int_0^\infty dl_1 \textrm{e}^{-\Sigma\left
(l_1-2^\zeta\frac{l_j+l_{j+1}}{2}\right)^2}\Bigg[ W(2^\zeta l_j',l_1)
\nonumber\\ + W(l_1,2^\zeta l_{j+1}') \Bigg]=W'_1(l_j',l_{j+1}')+
W'_2(l_j',l_{j+1}') 
\label{RGbindpot}
\end{eqnarray} 
The first term $W_1'$ corresponds to the contribution of the hard wall to the
renormalized binding potential, and $W_2'$ corresponds to the 
contribution of the original binding potential $W$. The hard wall contribution
can be evaluated exactly as:
\begin{eqnarray}
W'_1(l_j',l_{j+1}')=\frac{1}{2}\textrm{erfc }\left[\sqrt{\frac{2^{2\zeta} 
\Sigma}{\pi}}\left(\frac{l_j'+l_{j+1}'}{2}\right)
\right]\nonumber\\ \approx \frac{\exp\left[-\frac{2^{2\zeta}\Sigma}{\pi}
\left(\frac{l_j'+l_{j+1}'}{2}\right)^2\right]}
{\sqrt{2^{2\zeta}\Sigma}(l_j'+l_{j+1}')} \quad l_j'+l_{j+1}'\to \infty
\label{renw1}
\end{eqnarray}
For $W_2'$, we take into account the long distance behavior of $W$. After some
algebra, the leading order of $W_2'(l_j',l_{j+1}')$ can be written as:
\begin{equation}
W_2'(l_j',l_{j+1}')\approx -\frac{2^{1-\zeta p} A}{(\bar l_j')^p}
\int_{-s_0/2}^\infty ds 
\frac{\textrm{e}^{-s^2}/\sqrt{\pi}}
{\left[1+s/s_0\right]^p}
\label{renw2}
\end{equation}
where $\bar l_j'\equiv (l_j'+l_{j+1}')/2$ and $s_0\equiv \sqrt{\pi}
2^{\zeta+1}\bar l_j'$. As $\bar l_j'\to \infty$, the integral tends to 1
and we can see that $W'\sim W'_2 \sim A'/(\bar l_j')^p$, where $A'=A b^{1-\zeta
p}$. It is interesting to note that this result is also obtained by the 
following scaling argument for the binding potential:
\begin{equation}
\frac{A}{l^p}\to \frac{A'}{(l')^p}=b\frac{A}{l^p}=b\frac{A}{(b^\zeta l')^p}=
\frac{Ab^{1-\zeta p}}{(l')^p}
\label{heuristic}
\end{equation}
where we have taken into account that the binding potential is a free energy
per $x$ unit length. 

This analysis shows that the RG procedure leaves invariant the functional 
dependence of the asymptotic behavior of the binding potential. 
Two regimes can be identified. If $p>1/\zeta$, the binding potential 
strength decreases in each RG step. For $p<1/\zeta$  
the binding potential strength grows in each RG step. Finally, the
marginal case $p=1/\zeta$ corresponds to  
the leading asymptotic behavior remaining invariant. For $\zeta=1/2$, which
is the relevant value for planar wetting phenomena, this behavior leads to
the existence of two and three fluctuation regimes for complete and
critical wetting, respectively \cite{Fisher}. 
Note that this choice of $\zeta$ keeps the relevant microscopic scale 
$\xi_b\sim \Sigma^{-1}$ invariant. The analysis of the critical wetting 
from this RG approach can be found in Refs. \cite{Huse,Lipowsky,Spohn}.

A special class of effective hamiltonians are the following:
\begin{equation}
\exp[-\beta {\cal \tilde H}(l_i,l_{i+1})]=Z_\pi^{\Sigma,W}[l_i,l_{i+1};1]
\textrm{e}^{-\frac{\Sigma \theta^2}{2}}
\label{specialham}
\end{equation}
where $Z_\pi^{\Sigma,W}[l_i,l_{i+1};1]$ is the partition function 
Eq. (\ref{propagator}) with $x=a\equiv 1$ for arbitrary values of the
interfacial stiffness and binding potential $W(l)$.  
Such Hamiltonians can be regarded as those which are generated after one
iteration of the renormalization group provided that
$b$ very large. Indeed the fixed points found in Refs. 
\cite{Huse,Lipowsky,Spohn} belong to this class. 
Taking into account the properties of the path integrals, 
the renormalized potential after a RG step Eq. (\ref{RGtrans2}) can 
be written as:
\begin{equation}
\exp[-\beta {\cal \tilde H'}(l_j',l_{j+1}')]=b^\zeta 
Z_\pi^{\Sigma,W}[b^\zeta l_j',b^\zeta l_{j+1}';b]\textrm{e}^{-\frac{b \Sigma 
\theta^2}{2}}
\label{RGtrans4}
\end{equation}
For contact binding potentials Eq. (\ref{propagator}) and Kratzer potentials 
Eq. (\ref{laplacecoulomb}), it can be checked that 
Eq. (\ref{RGtrans4}) corresponds to an effective hamiltonian of the same 
form as the original since:
\begin{equation}
b^\zeta Z_\pi^{\Sigma,W}[b^\zeta l_j',b^\zeta l_{j+1}';b]=
Z_\pi^{\Sigma',W'}[l_j',l_{j+1}';1]
\label{RGtrans5}
\end{equation}
where $\Sigma$ and $\theta$ are transformed via Eq. (\ref{RGtrans3}) to 
$\Sigma'$ and $\theta'$. For Kratzer potentials $w$ must change as
$\Sigma^{-1}$, i.e. $w'=wb^{1-2\zeta}=w/b$ in order to preserve the 
invariance of the leading order behavior under renormalization.

Finally to finish our discussion of the RG for planar critical wetting
phenomena, we note that the $1-$point distribution function renormalizes
as:
\begin{equation}
P_\pi'(l_0')=b^\zeta P_\pi(b^\zeta l_0')
\label{RG1point}
\end{equation}

We will require this result later. Returning to our discussion about 
the RG in the wedge geometry, we need
to provide the transformation rules for $\alpha$ and $\beta f_W$. We will
assume that $\alpha$ changes as $\theta$:
\begin{equation}
\alpha \to \alpha'=\alpha b^{1-\zeta}
\label{RGtrans6}
\end{equation}
In order to obtain the transformation rule for $\beta f_W$, we consider how 
the $1-$point midpoint wedge probability distribution function renormalizes:
\begin{eqnarray}
P_w'(l_0',0)\equiv P_\pi'(l_0')\textrm{e}^{2\Sigma'\alpha' l_0'+ (\beta f_w)'}
\nonumber\\=b^\zeta P_\pi (b^\zeta l_0') \textrm{e}^{2\Sigma\alpha b^\zeta 
l_0'+ (\beta f_w)'} = b^\zeta P_w(b^\zeta l_0',0)
\label{RG1point2}
\end{eqnarray}
implying that $\beta f_w$ remains invariant:
\begin{equation}
\beta f_w \to (\beta f_w)'=\beta f_w
\label{RGtrans7}
\end{equation}
Finally, we note that if we change the effective hamiltonian by:
\begin{equation}
{\beta \cal H}(l_i,l_{i+1})={\beta \cal \tilde H}(l_i,l_{i+1})\pm [f(l_{i+1})-
f(l_{i})]
\label{newRG}
\end{equation}
where the sign is positive for $i\ge 0$ and negative for $i<0$, the 
probability of an interfacial configuration is now:
\begin{equation}
P_w(\{l_i\})=
\textrm{e}^{2\Sigma \alpha l_0+2f(l_0)+\beta f_W}\prod_{i=-\infty}^{\infty} 
\textrm{e}^{-\beta {\cal H}(l_i,l_{i+1})}
\label{probdisc2}
\end{equation}
The renormalization of the hamiltonian Eq. (\ref{RGtrans2}) is valid provided:
\begin{equation}
f'(l_i')=f(b^\zeta l_i')+C'
\label{newRG2}
\end{equation}
Note that any function $f(l_i)$ (unless it is a constant) breaks the
invariance of ${\beta \cal H}$ under exchange of its arguments and 
consequently introduces a directionality in the $x$ axis. 
This is perfectly sensible
in the wedge geometry, but is not admissible for the planar substrate, where
$-x$ is completely equivalent to $x$. A convenient choice for $f(l_i)$ is:
\begin{equation}
f(l_i)=\frac{1}{2}\ln P_\pi(l_i)
\label{fchoice}
\end{equation}
where $P_\pi(l_i)$ is the $1-$point probability distribution function
in the planar geometry, and the condition Eq. (\ref{fchoice}) is verified
due to Eq. (\ref{RG1point}). It is not difficult to see that 
$\exp(-{\beta \cal H})$ is the $2-$point conditional probability distribution 
function. Taking into account this fact and Eq. (\ref{RG1point2}), it is clear
that Eq. (\ref{probdisc2}) is exactly the same as Eq. (\ref{npointcorr2}).

The one remaining issue to be decided is the relevant value of the
exponent $\zeta$. We choose $\zeta=1$, so that the wedge tilt angle $\alpha$
and contact angle $\theta$ remain invariant in each step of the RG.

The procedure is now standard. The RG flow trajectories are constrained to the 
$\theta-$constant hypersurfaces in functional space.  
Since we know that the filling transition occurs for $\theta=\alpha$, we first
check that this situation corresponds to the critical manifold. 
Instead of considering an arbitrary potential, we choose as initial effective 
hamiltonians those of the form given by Eq. (\ref{specialham}), in particular 
with the partition function corresponding to contact binding potentials and 
those of Kratzer form. When the number of RG steps 
$n\to \infty$, the probability distribution of an interfacial configuration 
converges to a fixed point of the form:
\begin{equation}
P_w^*(\{l_i\})\propto l_0^{2\phi}\textrm{e}^{(\beta f_W)^*} 
\prod_{i=-\infty}^{\infty}\textrm{e}^{-{\beta \cal H^*}(l_i,l_{i+1})}
\label{fixedpoint}
\end{equation}
where $(\beta f_W)^*=-\infty$, and $\beta \cal H^*$ is defined as:
\begin{equation}
\textrm{e}^{-{\beta \cal H^*}(l_i,l_{i+1})}=
\begin{cases}
\delta(l_{i+1}-l_i+\alpha)H(l_i-\alpha)\\
+\delta(l_i)H(\alpha-l_i) & i\ge 0\\
\\
\delta(l_i-l_{i+1}+\alpha)H(l_{i+1}-\alpha)\\
+\delta(l_{i+1})H(\alpha-l_{i+1}) & i < 0
\end{cases}
\label{fixedpoint2}
\end{equation}
It is straightforward to see that Eq. (\ref{fixedpoint2}) is a fixed 
point of Eq. 
(\ref{RGtrans2}). Obviously, since $(\beta f_W)^*=-\infty$, the probability 
distribution for any interfacial configuration is zero. Nevertheless, we will 
consider Eq. (\ref{fixedpoint}) as formally different from zero. The basin of 
attraction for the $\phi=0$ case is expected to be those hamiltonians with
binding potentials that decay faster than $1/l$ (corresponding to
the filling fluctuation regime).
Hamiltonians that have a binding potential with a leading order 
$-\phi \alpha/l$ will be attracted to the fixed point Eq. (\ref{fixedpoint}) 
with the same value of $\phi$ for $\theta=\alpha$ (marginal case). Finally, 
if $lW(l)$ diverges as $l\to \infty$, there is no fixed point and 
the filling transition is mean-field-like. 

If $\theta \ne \alpha$, no matter how small $|\theta-\alpha|$ is, the RG
flow drives the hamiltonian away from the critical manifold. 
For $\theta > 0$, $\exp(-{\beta \cal H})$ converges to a fixed point expression
Eq. (\ref{fixedpoint2}) with $\alpha$ replaced by $\theta$. Even though
this new fixed point for the conditional probability is different from the 
the critical fixed point, the flow for the $2-$point conditional probability
distribution function will remain close (in some functional sense) to the 
critical one if $|\theta-\alpha|\ll \alpha$. Consequently, there is no 
relevant field associated with ${\beta \cal H}$. 

On the other hand, the wedge midpoint $1-$point probability distribution 
function behaves differently. We considered the same initial effective 
hamiltonians as for the $\theta=\alpha$ case. If $\theta < \alpha$, the 
distribution has an unphysical exponential growth with $l_0$ as 
$\exp(\Sigma |\theta-\alpha| l_0)$. The exponential term grows in each RG 
step, driving the probability distribution function to infinity (however, 
$\beta f_w=-\infty$, so the ``real'' probability of any interfacial 
configuration is zero). The attractor at infinity can be regarded as the 
complete filling fixed point. 
For $\theta > \alpha$, the wedge midpoint $1-$point probability distribution 
function becomes more and more peaked around zero as $\Sigma \to \infty$, 
converging to the low temperature fixed point:
\begin{equation}
P_w^{LT}(\{l_i\})=\prod_{i=-\infty}^\infty \delta(l_i)
\label{LTfixedpoint}
\end{equation}
These results imply that there is a relevant field (in the RG sense) 
associated with $P_w(l_0,0)$. Actually, the only other relevant operator is 
$h\propto \mu_c-\mu$, where $\mu$ is the chemical potential and $\mu_c$ 
the value at gas-liquid coexistence. 

Recall that the critical exponents defined at coexistence ($\mu=\mu_c$):
\begin{eqnarray}
\langle l(0)\rangle_w \sim t^{-\beta_w}\quad ; \quad 
\xi_\perp(0)\sim t^{-\nu_\perp}\nonumber\\
\xi_x \sim t^{-\nu_x}\quad ;\quad \beta f_W \sim t^{2-\alpha_w}
\label{critexp}
\end{eqnarray}
where $t=T_f-T$ and $T_f$ is the critical filling temperature. 
Close to a critical 
filling fixed point (for any $\phi$), all the relevant scale lengths 
$\Sigma^{-1}$, $\langle l(0)\rangle_w$, $\xi_x$, $\xi_\perp(0)$, 
etc. are reduced by a factor $1/b$. To extract the dependence on $t$ 
we need to know the largest eigenvalue of the linearized RG flow close 
to the critical fixed point. We argued above that this eigenvalue is 
associated to the transformation of the $x=0$ term of the hamiltonian 
${\beta \cal H}_0$. We again considered the special hamiltonians 
Eq. (\ref{specialham}) for contact and Kratzer binding potentials. The  
expression of ${\beta \cal H}_0$ for these potentials is:
\begin{equation}
{\beta \cal H}_0=2\Sigma (\theta -\alpha) l_0 - 2\phi \ln l_0 + C
\label{betah0}
\end{equation}
Taking into account how ${\beta \cal H}_0$ renormalizes, and its expression at
the critical fixed point (see Eq. (\ref{fixedpoint})), it is clear that the
largest eigenvalue is $b$, and its associated eigenfunction is proportional
to $\Sigma (\theta -\alpha) l_0$. As a consequence, the critical exponents 
$\nu_\perp=\nu_x=\beta_w=1$ for both the filling fluctuation and marginal
regimes. 

To get the leading singularity for the wedge excess free energy is more 
complicated. As $\beta f_w$ remains invariant under the RG transformations, 
$2-\alpha_w=0$. However, this value does not rule out a logarithmic 
divergence in $T_w-T$. In order to get such a dependence, we use the 
following thermodynamic relationship:
\begin{equation}
\left(\frac{\partial (\beta f_w)}{\partial \alpha}\right)_{\Sigma,\theta}=
-2\Sigma \langle l(0) \rangle_w
\label{thermodyn}
\end{equation}
where the derivative is made without changing $\Sigma$ or any characteristic 
of the binding potential (in particular the contact angle). 
We can rewrite Eq. (\ref{thermodyn}) as:
\begin{equation}
\left(\frac{\partial (\beta f_w)}{\partial 
\Sigma(\theta-\alpha)}\right)_{\Sigma,\theta}=2\langle l(0) \rangle_w
\label{thermodyn2}
\end{equation}
After a renormalization step, we know that, even when $\Sigma$ and the binding
potential have changed, this derivative changes as $1/b$ since $\beta f_w$, 
$\theta$ and $\alpha$ are invariant, and $\Sigma^{-1}$ and $\langle l(0) 
\rangle_w$ decrease by a factor $1/b$. If we regard 
$\Sigma (\theta-\alpha)$ ($\propto \Sigma$, 
and recall that $\theta-\alpha$ is invariant under a RG step) as proportional
to $t$ when the flow is close to the critical fixed point, 
Eq. (\ref{thermodyn2}) implies that $\beta f_w$ diverges logarithmically 
as $T\to T_f$. Furthermore, $\beta f_w$ is invariant under renormalization 
and must vanish as $\alpha \to 0$. Consequently close to the filling transition
it must be proportional dependence to $\ln(\theta-\alpha)-\ln(\theta)$. 

The critical exponents obtained for the critical filling transition are in 
complete agreement with exact calculations and scaling arguments 
\cite{Parry3}.  

\section{Conclusions \label{sec7}}

The structure of the gas-liquid interface bound at a 2D wedge and close to the 
filling transition has been studied using exact transfer-matrix methods. 
These calculations show the emergence of a new correlation length $\xi_F$
sufficiently close to the phase boundary ($\alpha<\theta<2\alpha$).
The explicit transfer matrix results for correlation functions and
interfacial roughness completely support the breather mode picture of 
fluctuation effects. The same picture also emerges of a renormalization
group approach that leaves the wedge geometry invariant. 
The fact that the only relevant fluctuations are those that translate 
the midpoint interfacial height leads to a simple relationship 
between the interfacial structure and midpoint height probability
distribution function in the scaling limit.
The extension of this approach to higher dimensions and/or 
different geometries would be very interesting and further work is being 
carried out in that direction.

Finally it would be instructive to understand the physical origin of the 
covariance relationship Eq. (\ref{covariance}). In the filling fluctuation 
regime we found that covariance can be inferred from the existence of the 
differential relation Eq. (\ref{fieldeq1}) and some regularity conditions. 
The very existence of such a field equation is itself indicative that 
some unknown symmetry relates wetting and filling transitions. 
Further work is required to elucidate whether any such hidden symmetry exists.

\acknowledgments
The authors would like to thank Dr. M. M\"uller for providing us the Ising
model simulation data from Ref. \cite{Albano}. J.M.R.-E. and M.J.G. 
acknowledge financial support from Secretar\'{\i}a de Estado de Educaci\'on 
y Universidades (Spain), co-financed by the European Social Fund, and 
EPSRC (UK), respectively.


\begin{thebibliography}{199}
\bibitem{Rejmer} K. Rejmer, S. Dietrich and M. Napiorkowski, Phys. Rev. E 
{\bf 60}, 4027 (1999).
\bibitem{Parry1} A. O. Parry, C. Rasc\'on and A. J. Wood, Phys. Rev. Lett. 
{\bf 83}, 5535 (1999).
\bibitem{Parry4} A. O. Parry, C. Rasc\'on and A. J. Wood, Phys. Rev. Lett. 
{\bf 85}, 345 (2000).
\bibitem{Parry2} A. O. Parry, A. J. Wood and C. Rasc\'on, J. Phys.: Condens.
Matter {\bf 13}, 4591 (2001).
\bibitem{Parry3} A. O. Parry, M. J. Greenall and A. J. Wood, J. Phys.: Condens.
Matter {\bf 14}, 1169 (2002).
\bibitem{Concus} P. Concus and R. Finn, Proc. Natl. Acad. Sci. USA {\bf 63}, 
292 (1969).
\bibitem{Pomeau} Y. Pomeau, J. Colloid Interface Sci. {\bf 113}, 5 (1986).
\bibitem{Hauge} E. H. Hauge, Phys. Rev. A {\bf 46}, 4994 (1992).
\bibitem{Abraham1} D. B. Abraham, A. O. Parry and A. J. Wood, Europhys.
Lett. {\bf 60}, 106 (2002).
\bibitem{Abraham2} D. B. Abraham and A. Maciolek, Phys. Rev. Lett. {\bf 89},
286101 (2002). 
\bibitem{Albano} E. V. Albano, A. De Virgiliis, M. M\"uller and K. Binder,
J. Phys.: Condens. Matter {\bf 15}, 333 (2003).
\bibitem{Wood} A. J. Wood and A. O. Parry, J. Phys. A: Math. Gen. {\bf 34}, L5
(2001).
\bibitem{Burkhardt} T. W. Burkhardt, Phys. Rev. B {\bf 40}, 6987 (1989) and
references therein.
\bibitem{Grosche} C. Grosche and F. Steiner, \emph{Handbook of Feynman
Path Integrals} (Springer, Berlin, 1998).
\bibitem{Greenall} M. J. Greenall, A. O. Parry and J. M. Romero-Enrique,
in preparation.
\bibitem{book} \emph{Tables of integral transforms}, vol. 1, edited
by A. Erd\'elyi (McGraw-Hill, New York, 1954).
\bibitem{Huse} D. A. Huse, Phys. Rev. Lett. {\bf 58}, 176 (1987).
\bibitem{Lipowsky} F. J\"ulicher, R. Lipowsky and H. M\"uller-Krumbhaar,
Europhys. Lett. {\bf 11}, 657 (1990).
\bibitem{Spohn} H. Spohn, Europhys. Lett. {\bf 14}, 689 (1991).
\bibitem{Fisher} R. Lipowsky and M. E. Fisher, Phys. Rev. B {\bf 36}, 
2126 (1987).
\end{thebibliography}
\end{document}